\documentclass[reprint,superscriptaddress,nofootinbib,amsmath,amssymb,aps,pra]{revtex4-2}

\usepackage{graphicx}
\usepackage{dcolumn}
\usepackage{bm}
\usepackage{braket}
\usepackage{mathrsfs}
\usepackage{hyperref}
\hypersetup{
    colorlinks,%
    citecolor=blue,%
    filecolor=blue,%
    linkcolor=blue,%
   	urlcolor=blue,
   	linktoc=page
}
\usepackage{physics}
\usepackage{slashed}
\usepackage[normalem]{ulem}

\newcommand{\bulkneq}{\mathrel{\stackrel{\makebox[0pt]{\mbox{\normalfont\tiny ?}}}{\neq}}}

\newcommand{\lambdatot}{{\lambda_{\text{tot}}}}

\begin{document}
\title{Hidden sectors of Chern-Simons matter theories and Exact Holography}
\author{Sachin Jain}
\author{Dhruva K.S}
\affiliation{Indian Institute of Science Education and Research,\\
 Dr Homi Bhabha Road, Pashan, Pune, India}
\author{Evgeny Skvortsov}
\affiliation{Service de Physique de l'Univers, Champs et Gravitation, \\ Universit\'e de Mons, 20 place du Parc, 7000 Mons,  Belgium}

\begin{abstract}
{
Nontrivial conformal field theories are rare. Motivated by the recent construction of a local higher-spin gravity in $AdS_4$ --- Chiral higher-spin gravity --- we begin to identify its AdS/CFT dual theory that has to be a closed subsector of Chern-Simons matter theories, thereby shedding more light on this class of CFTs and the dualities that relate them. We show that at the level of two- and three-point functions the sub-CFT can be defined as a limit of Chern-Simons matter theories where only correlators with the total positive (or negative) helicity are kept. We also discuss extension to higher orders and list a number of nontrivial predictions from the bulk. 
}
\end{abstract}
  \maketitle  
\newpage
\vspace*{-5mm}
\section{Introduction}

AdS/CFT duality \cite{Maldacena:1997re, Witten:1998qj, Gubser:1998bc} was born inside string theory and extended later with many examples that are not necessarily easy to embed into string theory. Most generally, AdS/CFT is the relation between theories of quantum gravity with negative cosmological constant and CFTs. Even more generally bulk theories can be dual to ensembles of CFTs. 

AdS/CFT is often applied as a ``knowledge transfer'' tool to deduce some nontrivial properties on one side of the duality from another one. One can also combine pieces of information coming from both sides of duality. Nevertheless,
for many reasons it would be important to have an example of a simple AdS/CFT duality where both sides are easy to define independently of each other and are simple enough to compute any relevant observable as a matter of principle. Ideally, this should lead to a direct proof of the holographic duality in such cases, which should shed more light on the mechanism of AdS/CFT and on the quantum gravity problem. 

It has been surprisingly hard to establish such simple AdS/CFT pairs, if we consider the canonical stringy dualities to be beyond technical capabilities to explore them in the whole space of couplings. For example, the supergravity limit where string theory is tractable corresponds to the regime of classical tensionful strings. Even putting the string loops aside, the opposite regime of tensionless strings is surprisingly hard to understand. In the canonical example of Type-IIB strings on $AdS_5\times S^5$ the tensionless limit corresponds to the weakly coupled $\mathcal{N}=4$ SYM \cite{HaggiMani:2000ru,Sundborg:2000wp,Sezgin:2001yf,Sezgin:2002rt,Beisert:2003te,Beisert:2004di}, i.e. a simple regime on the CFT side, but it has so far been difficult to take the tensionless limit directly on the string side.\footnote{It has been possible to directly land on the tensionless limit worldsheet theory in the case of string theory on $AdS_3$, see e.g. \cite{Eberhardt:2018ouy}.} The tensionful limit, on the other hand, corresponds to the strongly coupled SYM and can only be understood thanks to the integrability, see e.g. \cite{Beisert:2010jr}, which still gives access to a limited set of observables at present and is also hard to extend beyond the large-$N$ limit.

A natural idea to find tractable AdS/CFT pairs is to look for CFTs simpler/smaller than SYM and for bulk theories that are simpler than string theory. Among the latter, it is hard to find potentially UV-complete theories since it is related to the quantum gravity problem\footnote{It is possible to consider examples without a dynamical graviton, e.g. just the $\phi^4$ theory in $AdS_4$, thereby avoiding the quantum gravity problem. On the CFT side, one has to look for CFTs without the stress-tensor, which is not impossible, e.g. the long-range Ising model. }. Some examples that have extensively been studied over the years include $3d$-gravity as Chern-Simons theory, SYK, and JT models. Here, higher-spin gravities could be of some help. On the CFT side, the smallest possible conformal field theories are $3d$ vector models, in particular the $O(N)$ vector model, which describes second order phase transitions of many physical systems, e.g. the Ising critical point.

The space of $3d$ vector models can be substantially enriched by gauging some global symmetries therein and adding the Chern-Simons term, which results in Chern-Simons matter theories. One can also add supersymmetry and pass to models with bi-fundamental matter that extends to ABJ(M) theories \cite{Aharony:2008ug,Aharony:2008gk}. Remarkably, Chern-Simons matter theories are related to each other by a web of dualities, see e.g. \cite{Giombi:2011kc,Aharony:2011jz,Maldacena:2012sf, Aharony:2012nh,Aharony:2015mjs,Karch:2016sxi,Seiberg:2016gmd}, of which, perhaps, the most remarkable is the $3d$ bosonization duality \cite{Giombi:2011kc,Aharony:2011jz,Aharony:2012nh} since it remains to be nontrivial in the large-$N$ limit. 

Since (Chern-Simons) vector models are the smallest $3d$ CFTs it is natural to ask how the AdS/CFT dual thereof can look like \cite{Klebanov:2002ja,Sezgin:2003pt, Leigh:2003gk}. In the large-$N$ limit, it is easy to see the signs of an infinite-dimensional extension of conformal symmetry, called higher-spin symmetry. It is immediately visible in free vector models. One manifestation of the higher-spin symmetry is the presence of infinitely many conserved higher-spin tensors, $J_{a_1...a_s}$, of which the stress tensor is a particular member. By the standard AdS/CFT dictionary a spin-$s$ conserved tensor is dual to a gauge/massless spin-$s$ field on the AdS side. Therefore, the dual theory must feature an infinite multiplet of massless fields with spins from $0$ to $\infty$. Theories with massless higher-spin fields had been studied long before AdS/CFT correspondence and are usually called higher-spin gravities, see e.g. \cite{Bekaert:2022poo} for a review and e.g. \cite{Flato:1978qz,Bengtsson:1983pg,Berends:1984rq,Fradkin:1986ka,Fradkin:1987ks} for the first results in this direction. 

Many properties of higher-spin gravities can be deduced before even constructing a theory. A pro example of AdS/CFT correspondence is the Flato-Fronsdal theorem \cite{Flato:1978qz} that, in the modern language, states that the single trace operators in the free vector model are dual to the spectrum built from massless fields of all spins. However, constructing interactions faced a number of difficulties, the main being that higher-spin symmetry mixes spins and derivatives, thereby, rendering any generic higher-spin gravity too non-local, see e.g. \cite{Bekaert:2010hp,Bekaert:2015tva,Maldacena:2015iua,Sleight:2017pcz,Ponomarev:2017nrr,Ponomarev:2017qab,Neiman:2023orj}. There are very few exceptions: topological $3d$ higher-spin gravities \cite{Blencowe:1988gj,Bergshoeff:1989ns,Campoleoni:2010zq,Henneaux:2010xg,Pope:1989vj,Fradkin:1989xt,Grigoriev:2019xmp} and conformal higher-spin gravity \cite{Segal:2002gd,Tseytlin:2002gz,Bekaert:2010ky}. The former do not have propagating degrees of freedom and the latter extends conformal gravity. The AdS/CFT itself implies that the holographic duals of vector models do not obey the usual locality assumptions in field theory \cite{Bekaert:2015tva,Maldacena:2015iua,Sleight:2017pcz,Ponomarev:2017qab} and, hence, cannot be constructed by following, say, the Noether procedure. Nevertheless, certain structures that are stable under nonlocal field-redefinitions can be formulated as an $L_\infty$-algebra or as formally consistent classical equations of motion \cite{Vasiliev:1990en,Prokushkin:1998bq,Vasiliev:2003ev,Sagnotti:2005ns,Bonezzi:2016ttk,Bekaert:2017bpy,Arias:2017bvi,Grigoriev:2018wrx,Sharapov:2019vyd} and the dual of free/critical vector models can, in some sense, be reconstructed \cite{deMelloKoch:2018ivk,Aharony:2020omh}. The former leaves infinitely many coefficients unfixed and does not allow for systematic holographic calculations, while the latter does not give an independent definition of the bulk theory and cannot at present be extended to Chern-Simons matter theories. 

Therefore, it turns out that simple CFTs, like free vector models, do not have simple holographic duals in the sense that the bulk theories have to be too nonlocal for the standard field theory rules to apply and new techniques are yet to be developed to construct and deal with them.

A new impulse into this puzzle came from the flat space, an unusual place for AdS/CFT duality, when it was shown that there exists a local theory with the required spectrum \cite{Metsaev:1991nb,Metsaev:1991mt,Ponomarev:2016lrm} --- chiral higher-spin gravity. It was conjectured already in \cite{Ponomarev:2016lrm} that there should exist a smooth deformation to $(A)dS_4$, which was constructed later in \cite{Metsaev:2018xip,Skvortsov:2018uru,Sharapov:2022wpz,Sharapov:2022awp}. The theory can be interpreted as a higher-spin extension of both self-dual Yang-Mills and self-dual gravity. It was shown to be at least one-loop finite \cite{Skvortsov:2018jea,Skvortsov:2020gpn}. It is also integrable at least in flat space \cite{Ponomarev:2017nrr}. Therefore, there exists a tame theory in $AdS_4$ with the right spectrum to be dual to Chern-Simons matter theories. However, it cannot be dual to the full Chern-Simons matter theory. For example, the chirality of interactions is such that the sum of helicities must be positive (or negative for the anti-chiral theory) at each cubic vertex.

Given there is a well-defined (anti-)chiral higher-spin gravity in $AdS_4$ we can apply the $AdS/CFT$ correspondence to conclude that there should exist two closed hidden subsectors in Chern-Simons matter theories that are consistent on their own. \\
\begin{figure}
    \includegraphics[trim={1cm 0.9cm 1cm 0},clip,width=0.27\textwidth]{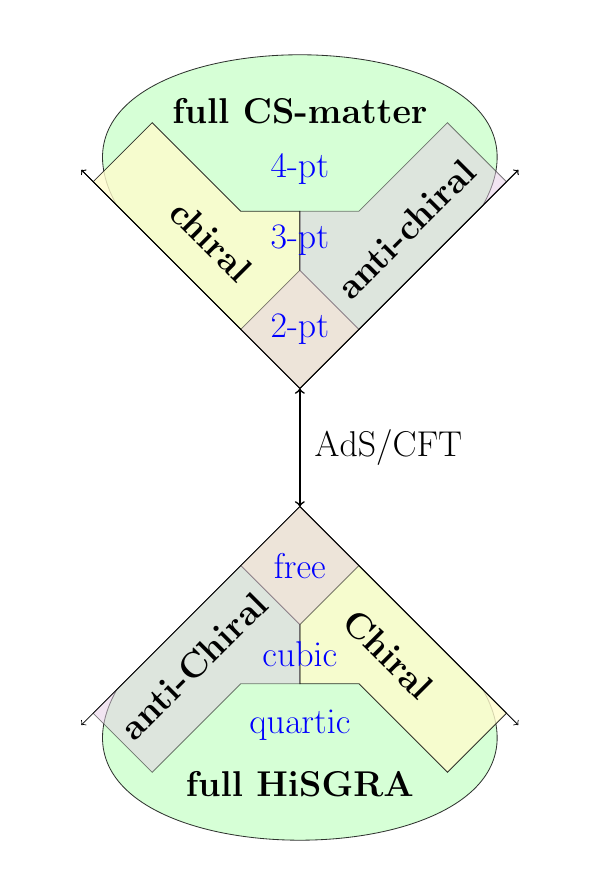}
    \caption{Chern-Simons vector models vs. higher-spin gravity: the  structure of AdS/CFT duality. The spectrum of fields/operators is the same (free theory in the bulk/$2$-point correlators on the boundary). (Anti)-chiral sectors together cover all cubic vertices/three-point functions. There are non-chiral parts at higher orders. }
    \label{fig:main}
    \end{figure}

Another simple consequence of there being an $AdS$ theory is the $3d$ bosonization at the level of $3$-point correlators \cite{Skvortsov:2018uru}. Indeed, chiral and anti-chiral interactions form a complete basis of cubic interactions and all the couplings are completely fixed in each of the sectors. Therefore, in order to bootstrap the $3$-point correlators in a theory with the same spectrum of higher-spin fields, one needs to see how to glue (anti-)chiral pieces together. The gluing depends on one phase-like parameter that gives the structure observed by Maldacena and Zhiboedov in \cite{Maldacena:2012sf}. Beyond $3$-point (anti-)chiral interactions do not cover everything possible, but see \cite{Skvortsov:2022wzo}. The structure of the duality is illustrated on the figure.\\

On the CFT side there has been a lot of progress in understanding the Chern-Simons matter theories in the large-$N$ limit, a significant part of which was based on the idea of the slightly-broken higher-spin symmetry \cite{Giombi:2011kc,Aharony:2011jz,Maldacena:2011jn,Maldacena:2012sf}. Several remarkable achievements in these theories include exact computations of the partition function \cite{Giombi:2011kc,Aharony:2011jz,Jain:2012qi,Aharony:2012ns,Jain:2013py,Jain:2013gza,Minwalla:2023esg}, understanding the Hilbert space structure \cite{Minwalla:2022sef} and the higher-spin spectrum \cite{GurAri:2012is,Takimi:2013zca,Giombi:2016zwa,Giombi:2017rhm,Charan:2017jyc,Jain:2019fja,Minwalla:2020ysu}, determination of the exact S matrix \cite{Jain:2014nza,Inbasekar:2015tsa,Mehta:2022lgq}, showing the existence of BCFW recursion relations and dual superconformal invariance \cite{Inbasekar:2017ieo,Inbasekar:2017sqp}.

 An important set of observables in these theories are the correlation functions of gauge invariant single trace operators, aka higher-spin currents. It was found in \cite{Giombi:2011rz} that three-point functions in general three-dimensional CFTs involving higher-spin currents contain three pieces: two of which are realized by the free bosonic and free fermionic theories and a third, which is parity violating and not realized by free fields. In \cite{Maldacena:2011jn,Maldacena:2012sf} it was found that if one imposes in addition, the Ward identities due to the slightly-broken higher-spin symmetry, the relative coefficients of these structures get determined in terms of a single parameter. This parameter was then related to the Chern-Simons matter theory 't Hooft coupling in \cite{Aharony:2012nh}. Since then, there have been many attempts to compute three- and four-point correlators in position space and Mellin space 
\cite{Chowdhury:2017vel,Sezgin:2017jgm,Skvortsov:2018uru,Yacoby:2018yvy,Chowdhury:2018uyv,Aharony:2018pjn,Gerasimenko:2021sxj,Scalea:2023dpw, Bedhotiya:2015uga,Turiaci:2018nua,Li:2019twz,Silva:2021ece,Kalloor:2019xjb,Trivko:2024thesis,Kukolj:2024yyo} and also in momentum space \cite{Jain:2020rmw,Jain:2020puw}. It was subsequently realized by working in momentum space, that the parity odd correlators can be obtained from the difference of the free fermionic and free bosonic correlators by the so-called \textit{epsilon} transform \cite{Jain:2021gwa,Jain:2021whr,Caron-Huot:2021kjy}. Further, by employing the technology of three-dimensional spinor-helicity variables, this relation becomes extremely simple \cite{Jain:2021vrv,Jain:2021wyn,Jain:2021qcl}. Spinor-helicity variables also revealed the anyonic nature of the correlators \cite{Gandhi:2021gwn}. Eventually, this culminated in the analysis of all four- and higher-point functions in the Chern-Simons matter theories \cite{Jain:2022ajd} using the slightly-broken higher-spin symmetry. Based on these results, the implications on the bulk dual of Chern-Simons matter theory were analyzed in \cite{Jain:2023juk} where it was found that there exists the possibility of a sub-sector of the theory that has a local bulk dual.

Since chiral higher-spin gravity is a perturbatively local field theory, one can directly compute holographic correlators to define the hidden subsectors of Chern-Simons matter theories. However, it would be important to identify the hidden sectors directly on the CFT side. This is what the present paper attempts to do. Crucial ingredients for our analysis are the epsilon transform and the usage of three-dimensional spinor-helicity variables which together manifest the anyonic nature of correlation functions. We shall show that taking appropriate limits of the 't Hooft coupling of the theory projects us into a chiral or anti-chiral sector at the level of three-point correlators. We also derive some simple constraints for the structure of higher-order correlators and loop corrections from the bulk. The loop corrections are shown to truncate at a finite order in $1/N$-expansion, which makes the latter convergent. All of this paves the way to constructing exact models of AdS/CFT correspondence.\\

\paragraph{Outline:} 
The paper is organized as follows: In section \ref{sec:CFT3corrfacts}, we first state and prove a number of essential facts about two and three point correlators in three dimensional CFTs. We then use these results to define the (anti-)chiral sub-CFT of Chern-Simons matter theories in section \ref{sec:ChiralHSfromCS}. In section \ref{sec:beyond}, we analyze some higher point functions and the anomalous dimensions of the currents in the (anti-)chiral limit. We then make some simple predictions from the bulk that are based on the known equations of motion and parts of the action of Chiral theory in section \ref{sec:holo}. Finally, in section \ref{sec:conc}, we end with future perspectives and conclusions. We also have nine appendices to elaborate on and complement the main text.

\section{Some generalities about correlators in $\text{CFT}_3$}\label{sec:CFT3corrfacts}

In this section, we shall discuss and derive some general facts about two and three-point correlators in three-dimensional CFT involving scalars and conserved currents. We work throughout in momentum space and in three-dimensional spinor helicity variables \cite{Bzowski:2013sza,Coriano:2013jba,Bzowski:2015pba,Bzowski:2017poo,Jain:2020rmw,Jain:2020puw,Jain:2021wyn,Jain:2021vrv,Jain:2021gwa,Jain:2021whr,Maldacena:2011nz}.  We provide a guide to our notation and conventions in appendix \ref{appendix:Notation}.
\subsection{Two-point functions}
The two-point function of a primary scalar operator $O_{\Delta}$ is fixed by the conformal ward identities up to an overall constant. In momentum space, it is given by the following expression \cite{Bzowski:2013sza},\footnote{In this equation and the ones to follow, we suppress the overall momentum conserving delta functions.}
\begin{align}
    \langle O_{\Delta}(p)O_{\Delta}(-p)\rangle=c_{\Delta}~p^{2\Delta-3}.
\end{align}
Two-point functions of conserved currents are similarly constrained and determined up to two constants, one of which multiplies the usual parity even expression while the other, multiplies a parity odd contact term \cite{Jain:2021gwa}. This is an artefact of three dimensions, due to the presence of the three indexed Levi-Civita symbol. For example, the two-point function of a spin-one conserved current is given by,\footnote{Given a spin $s$ current $J_s^{\mu_1\cdots\mu_s}(p_i)$, it is often convenient to contract it with the polarization vectors $z_{i\mu_1}\cdots z_{i\mu_s}$. These polarization vectors are null ($z_i^2=0$) and transverse ($z_i\cdot p_i=0$).}
\begin{align}\label{JJ2point}
    z_{1\mu}z_{2\nu}\langle J^{\mu}(p) J^{\nu}(-p)\rangle=c_{1,\text{even}} ~p~(z_1\cdot z_2)+c_{1,\text{odd}}~ \epsilon^{z_1 z_2 p}.
\end{align}
This expression is easily generalized to the spin $s$ case:
\footnotesize
\begin{align}\label{JsJs2point}
       &z_{1\mu_1}\cdots z_{1\mu_s}z_{2\nu_1}\cdots z_{2\nu_s}\langle J_s^{\mu_1\cdots\mu_s}(p) J_s^{\nu_1\cdots \nu_s}(-p)\rangle\notag\\&=c_{s,\text{even}}~p^{2s-1}~(z_1\cdot z_2)^{s}+c_{s,\text{odd}}~p^{2s-2}~\epsilon^{z_1 z_2 p}(z_1\cdot z_2)^{s-1}\notag\\
       &=z_{1\mu_1}\cdots z_{1\mu_s}z_{2\nu_1}\cdots z_{2\nu_s}\bigg(c_{s,\text{even}}\langle J_s^{\mu_1\cdots\mu_s}(p) J_s^{\nu_1\cdots \nu_s}(-p)\rangle_{\text{even}}\notag\\&\qquad\qquad\qquad~~~~~~~~~~~~~~~~+c_{s,\text{odd}}\langle J_s^{\mu_1\cdots\mu_s}(p) J_s^{\nu_1\cdots \nu_s}(-p)\rangle_{\text{odd}}\bigg)\,,
\end{align}
\normalsize
where we used the notation $\epsilon^{\mu\nu\rho}a_\mu b_\nu c_\rho:=\epsilon^{abc}$. It is interesting to note that the parity odd expression can be obtained from the parity even one via the \textit{epsilon} transform as follows \cite{Jain:2021gwa}:
\small
\begin{align}\label{2pointEPToddeven}
    &\langle J_s^{\mu_1\cdots \mu_s}(p)J_s^{\nu_1\cdots\nu_s}(-p)\rangle_{\text{odd}}\notag\\&=\frac{1}{p}{\epsilon^{ p \alpha(\mu_1}}\langle J_s^{\mu_2\cdots \mu_s)\alpha}(p)J_s^{\nu_1\cdots\nu_s}(-p)\rangle_{\text{even}}.
\end{align}
\normalsize
One may think that the $c_{s,\text{odd}}$ contribution can be removed via the addition of suitable counter-terms as it multiplies a contact term contribution\footnote{$ p^{2(s-1)}\epsilon^{z_1z_2p}(z_1\cdot z_2)^{s-1}$ is the Fourier transform of the following contact term:\\ $(z_1\cdot z_2)^{s-1}\epsilon^{z_1 z_2 \mu}\partial_{1\mu}\Box_1^{2(s-1)}\delta^3(x_1-x_2)$.}. This, however, is not completely true as there could exist a scheme-independent part of the parity odd two-point functions and hence no counter-terms can remove them. This is explicitly realized in Chern-Simons theory with matter \cite{Aharony:2012nh}. In fact, these contact terms will play a crucial role in section \ref{sec:ChiralHSfromCS} while defining the chiral limit of Chern-Simons matter theory. 

It turns out that \eqref{JsJs2point} takes an interesting form in the language of three-dimensional spinor-helicity variables. For a review of the same with our conventions please refer to appendix \ref{appendix:Notation}. In the two nonzero helicity configurations, i.e., $(-~-)$ and $(+~+)$, we have,
\begin{align}\label{JsJsinmmandpp}
    &\langle J_s^{-}(p)J_s^{-}(-p)\rangle\sim(c_{s,\text{even}}-i c_{s,\text{odd}})\frac{\langle 1 2\rangle^{2s}}{p},\notag\\&\langle J_s^{+}(p)J_s^{+}(-p)\rangle\sim(c_{s,\text{even}}+i c_{s,\text{odd}})\frac{\langle \Bar{1} \Bar{2}\rangle^{2s}}{p}.
\end{align}
We see that in spinor-helicity variables, the even and odd terms are identical up to a helicity (but not momentum) dependent factor. This factor of $\mp i$ can be understood by looking at the epsilon transform \eqref{2pointEPToddeven}. Effectively, the epsilon transform with respect to a conserved current $J_{s}$ becomes a simple multiplicative transformation in spinor-helicity variables.\footnote{Consider the $s=1$ case for instance with a negative helicity polarization.
\begin{align}
    z_{\mu}^{-}(\epsilon\cdot J_s)^{\mu}:=z_{\mu}^{-}\frac{\epsilon^{\mu p \alpha}}{p}J_{\alpha}.
\end{align}
By using the fact that $\epsilon^{\mu \nu \rho}=\frac{1}{2i}\Tr{\sigma^{\mu}\sigma^{\nu}\sigma^{\rho}}$ and $z_{\mu}^{-}=\frac{(\sigma_\mu)^{a}_b\lambda_a\lambda^b}{2p}, p^{\mu}=\frac{(\sigma^\mu)^{a}_b\lambda_a\Bar{\lambda}^b}{2}$, we find that,
\begin{align}
    z_{\mu}^{-}(\epsilon\cdot J_s)^{\mu}=-iz^{\alpha-}J_{\alpha}=-i J^{-}.
\end{align}
For the case when the polarization has a positive helicity, we obtain a $+i$ on the RHS instead of the $-i$. Since the polarization vectors for higher-spins are formed out of products of the spin one polarization, the same results hold there. In short, if the polarization has helicity $h$ we obtain a $\text{sign}(h)i$ factor while performing the epsilon transform.}

In section \ref{sec:ChiralHSfromCS}, we will see that it is the structure of \eqref{JsJsinmmandpp} that will enable us to define the chiral limit of Chern-Simons matter theory by appropriately relating the odd coefficient that appears there to the even one.
\subsection{ Three-point functions}
Three-point correlators of conserved currents can be grouped into two broad categories. Given a three-point function $\langle J_{s_1}J_{s_2}J_{s_3}\rangle$ such that $s_i+s_j\ge s_k~\forall i,j,k\in\{1,2,3\}$, we say that the correlator is inside the (spin) triangle. If this inequality is violated, we say that it is outisde the (spin) triangle. In this subsection, 
we first review some general facts about correlators both inside and outside the triangle. We then proceed to derive some new results which will play an important role in section \ref{sec:ChiralHSfromCS}.
\subsubsection{\textbf{Three-point functions inside the triangle}}\label{3ptouttriangle}
The most general three-point function that is inside the triangle takes the following form \cite{Giombi:2011rz}:\footnote{\label{ft:AA}There could also be parity odd nonhomogeneous contributions but they are all contact terms. Since the FF and FB Ward-Takahashi identities are equal inside the triangle, we can form one parity odd nonhomogeneous structure by replacing the two-point function that appears there by its parity odd counterpart \cite{Jain:2021vrv}.}
\small
\begin{align}\label{3ptCFT}
    \langle J_{s_1}J_{s_2}J_{s_3}\rangle=n_f \langle J_{s_1}J_{s_2}J_{s_3}\rangle_{FF}+n_{odd}\langle J_{s_1}J_{s_2}J_{s_3}\rangle_{odd}+n_b \langle J_{s_1}J_{s_2}J_{s_3}\rangle_{FB}.
\end{align}
\normalsize
We have used the abbreviations $FF=$free fermion and $FB=$free boson that we shall employ throughout this paper.
It was found in \cite{Jain:2021gwa} that the odd part can be written in terms of the FF and FB results via an epsilon transformation.
\small
\begin{align}\label{oddepT}
    \langle J_{s_1}J_{s_2}J_{s_3}\rangle_{odd}&= \langle \epsilon\cdot J_{s_1}J_{s_2}J_{s_3}\rangle_{FF-FB}\notag\\&=\langle  \epsilon\cdot J_{s_1}J_{s_2}J_{s_3}\rangle_{FF}-\langle  \epsilon\cdot J_{s_1}J_{s_2}J_{s_3}\rangle_{FB}.
\end{align}
\normalsize
 For example, consider \eqref{oddepT} for the stress tensor three point function. Keeping all indices explicit, it is given by,
\begin{align}
    \langle  T_{\mu \nu}(p_1)T_{\alpha\beta}(p_2)&T_{\gamma\theta}(p_3)\rangle_{\text{odd}}=\langle  \epsilon\cdot T_{\mu \nu}(p_1)T_{\alpha\beta}(p_2)T_{\gamma\theta}(p_3)\rangle_{FF-FB}\notag\\&=\frac{\epsilon_{ a b(\mu}p_{1a}}{p_1}\langle T_{\nu) b}(p_1)T_{\alpha\beta}(p_2)T_{\gamma\theta}(p_3)\rangle_{FF-FB}.
\end{align}
It is important to note that for correlators inside the triangle, one can obtain the odd correlator via the epsilon transform with respect to any of the currents in the $FF-FB$ correlator.\footnote{This is due to 
the fact that the $FF-FB$ correlators are non-zero only when all their helicities have the same sign as we shall see in \eqref{statement1}. Thus, the epsilon transform, which gives a factor of $i~\text{sign}(h)$ can be performed with respect to any of the operators.} We will see that the same cannot be said for correlators that violate the triangle inequality in the next subsection.\\

Further, we define the homogeneous and nonhomogeneous correlators \cite{Jain:2021gwa},\footnote{The homogeneous and nonhomogeneous correlators have nice bulk interpretations. The former comes due to higher derivative interactions whereas the latter comes due to the term that also gives rise to the propagators of the involved bulk fields.}:
\begin{align}\label{handnh}
    &\langle J_{s_1}J_{s_2}J_{s_3}\rangle_{h}=\frac{1}{2}\langle J_{s_1}J_{s_2}J_{s_3}\rangle_{FF-FB}\notag,\notag\\
    &\langle J_{s_1}J_{s_2}J_{s_3}\rangle_{nh}=\frac{1}{2}\langle J_{s_1}J_{s_2}J_{s_3}\rangle_{FF+FB},
\end{align}
where we used the notation $\langle\cdots\rangle_{FF\pm FB}=\langle \cdots\rangle_{FF}\pm\langle \cdots\rangle_{FB}$.
These quantities satisfy,
\begin{align}
    &p_{1\mu}\langle J_{s_1}^{\mu\cdots }J_{s_2}J_{s_3}\rangle_{h}=0,\notag\\
    &p_{1\mu}\langle J_{s_1}^{\mu\cdots}J_{s_2}J_{s_3}\rangle_{nh}=\text{Ward-Takahashi identity terms}.
\end{align}
Since the Ward-Takahashi identity is made out of two-point functions, the coefficient of the nonhomogeneous correlator is that of the two-point function.

Now, using \eqref{oddepT} and \eqref{handnh}, \eqref{3ptCFT} can be written as,
\begin{align}\label{3ptCFThandnh}
    \langle J_{s_1}J_{s_2}J_{s_3}\rangle&=(n_f+n_b)\langle J_{s_1}J_{s_2}J_{s_3}\rangle_{nh}+(n_f-n_b)\langle J_{s_1}J_{s_2}J_{s_3}\rangle_{h}\notag\\&+2n_{\text{odd}}~ \langle \epsilon\cdot J_{s_1}J_{s_2}J_{s_3}\rangle_{h}.
\end{align}
As mentioned just above this equation, the coefficient of the nonhomogeneous correlator viz $n_f+n_b$ is proportional to the two-point function coefficient.\footnote{For instance, consider the $\langle TT\rangle$ two-point function: $\langle TT\rangle\propto C_T$. We then have $\langle TTT\rangle_{nh}\propto C_T$ since $p_1\cdot\langle TTT\rangle\sim\langle TT\rangle$. This also implies in \eqref{3ptCFThandnh} that $n_f+n_b=C_T$.} $n_f-n_b$ and $n_{\text{odd}}$ on the other hand are truly three-point data.
Thus we see that in contrast to the three structures appearing in \eqref{3ptCFT}, only two appear in \eqref{3ptCFThandnh} thanks to the epsilon transformation that related the odd part to the $FF-FB$ (homogeneous) correlator. Further, these homogeneous and nonhomogeneous correlators obey many interesting properties. However, for the best way to see these properties, we must turn to the language of three-dimensional spinor-helicity variables as we did even for the two-point functions in \eqref{JsJsinmmandpp}. For instance, \eqref{3ptCFThandnh} in spinor-helicity variables takes the following form (where $h_i$ is the helicity of $J_{s_i}$) in spinor-helicity variables:
\begin{align}
     \langle J_{s_1}^{h_1}J_{s_2}^{h_2}J_{s_3}^{h_3}\rangle&=(n_f+n_b)\langle J_{s_1}^{h_1}J_{s_2}^{h_2}J_{s_3}^{h_3}\rangle_{nh}\notag\\&+(n_f-n_b\pm 2i n_{\text{odd}})\langle J_{s_1}^{h_1}J_{s_2}^{h_2}J_{s_3}^{h_3}\rangle_{h},
\end{align}
where the $\pm i$ multiplying $n_{\text{odd}}$ depends on the helicity configuration in question.

We shall now state some facts about the (non)homogeneous correlators and subsequently prove them.\\\\
\underline{\textbf{Statement}~1:} For correlators inside the triangle, the homogeneous parts, $ \langle J_{s_1}J_{s_2}J_{s_3}\rangle_{h}$ are only nonzero in the $(- - -)$ and $(+ + +)$ helicity configurations \cite{Jain:2021vrv}. It is important to note that the odd correlators in spinor-helicity variables are equal to the even homogeneous correlators up to factors of $\pm 2i$ that depend on the helicity\footnote{The fact that homogeneous correlators are dual to higher derivative interactions in the bulk supports this statement. Such interactions are nonzero only when the helicities of all three particles coincide. For example, consider $\langle TTT\rangle_h$. It can be obtained from the graviton $W^3$ (Weyl tensor cubed) interaction in the bulk which vanishes except when the gravitons all have the same helicity.}
\begin{align}\label{statement1}
    &\langle J_{s_1}^{-}J_{s_2}^{-}J_{s_3}^{-}\rangle_h\ne 0\,,\qquad \langle J_{s_1}^{+}J_{s_2}^{+}J_{s_3}^{+}\rangle_h\ne 0\,,\notag\\&\langle J_{s_1}^{h_1}J_{s_2}^{h_2}J_{s_3}^{h_3}\rangle_h=0~\text{for all other helicities}.
\end{align}
 \underline{\textbf{Statement}~2:} The nonhomogeneous correlators inside the triangle, $ \langle J_{s_1}J_{s_2}J_{s_3}\rangle_{nh}$ are only nonzero in the mixed helicity configurations, i.e. they are zero in the $(- - -)$ and $(+ + +)$ helicity configurations. We show that the nonhomogeneous correlators that appear in \cite{Jain:2021vrv} can be modified via the addition of contact terms to obey this statement\footnote{Nonhomogeneous correlators arise from the $s$ derivative interactions in the bulk which have support only if one of the particles has a different helicity than the other two. For instance consider $\langle JJJ\rangle_{nh}$ (non-Abelian currents). This correlator is dual to the usual Yang-Mills interaction in the bulk which is nonzero only in the $(--+)$ configuration and those obtained via exchanges or complex conjugation.}
 \begin{align}\label{statement2}
    &\langle J_{s_1}^{-}J_{s_2}^{-}J_{s_3}^{-}\rangle_{nh}=\langle J_{s_1}^{+}J_{s_2}^{+}J_{s_3}^{+}\rangle_{nh}=0\,,~~~\notag\\
    &\langle J_{s_1}^{h_1}J_{s_2}^{h_2}J_{s_3}^{h_3}\rangle_{nh}\ne 0~\text{for other (net nonzero) helicities}.
\end{align}
\underline{\textbf{Statement}~3:} The zero-helicity sectors (i.e. the sector where the net helicity of the correlator is zero) of the nonhomogeneous correlators is identically zero. Similar to what we discussed in the above statement, this statement too will be true modulo contact term contributions. Put another way, we will show that we can find contact terms to remove such contributions to the nonhomogeneous correlators\footnote{As the CFT correlators give rise to S-matrices in one higher dimension when the total energy goes to zero and zero-helicity three-point S-matrices are identically zero, the CFT correlator should be free of total energy poles. Thus, it is plausible (as we shall also show) that the zero-helicity CFT correlators can be removed via suitable contact terms. For a slightly more detailed discussion regarding this as well as an argument for statement-2 from the CFT-S Matrix correspondence, please refer to appendix \ref{appendix:CFTsmatrixstuff}.}
\begin{align}\label{statement3}
    \langle J_{s_1}^{h_1}J_{s_2}^{h_2}J_{s_3}^{h_3}\rangle=0~~\text{if $s_1h_1+s_2h_2+s_3h_3=0$}.
\end{align}
Let us now present a suggestive proof for statement 1, \eqref{statement1}: consistency of the momentum space correlator with the operator product expansion forces it to only have total energy poles \cite{Maldacena:2011nz}. The other type of poles which are of the form $p_i+p_j-p_k$, are inconsistent with the OPE. The momentum space expressions that give rise to the correct conformally invariant homogeneous correlators in the $(---)$ and $(+++)$ helicities were found in \cite{Jain:2021vrv}. These expressions also yield zero in the mixed helicity configurations. Thus, in principle, any other solution that gives rise to non-zero expressions in the mixed helicities will also ruin the correct $(---),(+++)$ answers. This is further solidified by the uniqueness of the momentum space solution: Other solutions (that are consistent with the OPE) simply cannot exist  except of course, modulo some contact term contributions.  Also, it is interesting to note that when the spin-triangle inequality is violated, there is no homogeneous solution in any helicity configuration consistent with the OPE \cite{Jain:2021whr}.


We now present a few examples to illustrate the validity of statements 2 \eqref{statement2} and 3 \eqref{statement3}. 
\subsection*{$\textbf{Example~1}: \langle JJJ\rangle$}
This correlator in the two independent helicity configurations takes the form \cite{Jain:2021vrv}:\footnote{In this example and the ones to follow, we focus on nonhomogeneous correlators that are parity even. Parity odd nonhomogeneous correlators are contact terms \cite{Jain:2021vrv} and hence we do not keep them in the analysis to follow.}
\begin{align}\label{JJJcorr}
    &\langle J^{-A}J^{-B}J^{-C}\rangle=f^{ABC}\big(\frac{c_{E}+ic_{O}}{E^3}+\frac{c_J}{p_1p_2p_3}\big)\langle 1 2\rangle\langle 2 3\rangle\langle 3 1\rangle\,,\notag\\
    &\langle J^{-A}J^{-B}J^{+C}\rangle=f^{ABC}\frac{c_J}{p_1p_2p_3}\big(1-\frac{2p_3}{E}\big)\langle 1 2\rangle\langle 2 \Bar{3}\rangle\langle \Bar{3}1\rangle\notag\,,\\
    &\text{where}~E=p_1+p_2+p_3\,,
\end{align}
where $c_{E}$ ($ic_{O}$) multiplies the parity even (parity odd) homogeneous contribution and
$c_J$ is the coefficient of the parity even nonhomogeneous term. The $(+++)$ and $(++-)$ configurations can be obtained from the above expression by complex conjugation (which replaces $\langle i j\rangle\to \langle \Bar{i}\Bar{j}\rangle$). Finally, the remaining helicities are obtained via exchanges of labels.\\

Looking at \eqref{JJJcorr}, we see that the homogeneous solution is present only in the $(---)$ and $(+++)$ helicities which is in accordance to statement $1$. However, it appears as though there is a nonhomogeneous contribution in the $(---)$ and $(+++)$ helicities which should not be there according to statement $2$ \eqref{statement2}. We shall show however, that via the addition of a suitable contact term, we can cancel these contributions. Indeed, consider the following contact term contribution:
\small
\begin{align}\label{JJJcontact}
        &\langle J^{A}J^{B}J^{C}\rangle_{\text{contact}}=\frac{2c_J f^{ABC}}{3p_1p_2p_3}\bigg((z_1\cdot z_2)(z_3\cdot p_1)(p_1+p_2)p_3+\notag\\
        &-(z_1\cdot z_3)(z_2\cdot p_1)p_2(p_1+p_3)+(z_1\cdot p_2)(z_2\cdot z_3)p_1(p_2+p_3)\bigg).
\end{align}
\normalsize
We then redefine the correlator via the addition of \eqref{JJJcontact} as follows:
    $\langle J^{A}J^{B}J^{C}\rangle_{new}=\langle J^{A}J^{B}J^{C}\rangle-\langle J^{A}J^{B}J^{C}\rangle_{\text{contact}}$.
This results in the following expression for $\langle JJJ\rangle_{new}$ in the two independent helicity configurations:
\begin{align}\label{JJJcorrnew}
    &\langle J^{-A}J^{-B}J^{-C}\rangle_{new}=f^{ABC}\big(\frac{c_E+i c_O}{E^3}\big)\langle 1 2\rangle\langle 2 3\rangle\langle 3 1\rangle,\notag\\
    &\langle J^{-A}J^{-B}J^{+C}\rangle_{new}=f^{ABC}\frac{2c_J}{3p_1p_2p_3}\big(1-\frac{3p_3}{E}\big)\langle 1 2\rangle\langle 2 \Bar{3}\rangle\langle \Bar{3}1\rangle,
\end{align}
proving that the nonhomogeneous contribution to $\langle JJJ\rangle$ can be tuned to zero in the $(- - -)$ and $(+++)$ helicities by the addition of the contact term \eqref{JJJcontact}.\footnote{The same is true in the $(+++)$ helicity as it is the complex conjugate of the $(---)$ helicity.} 
\subsection*{$\textbf{Example~2}:\langle TTT\rangle$}
Let us now consider the three-point function of the stress tensor. This correlator takes the following form in the two independent helicities \cite{Jain:2021vrv}:
\begin{widetext}
\begin{align}
    &\langle T^{-}T^{-}T^{-}\rangle=\bigg(\frac{(c_E+ic_O) p_1 p_2 p_3}{E^6}+c_T\frac{1}{(p_1p_2p_3)^2}\big(E^3-E(p_1p_2+p_2p_3+p_1p_3)-p_1p_2p_3\big)\bigg)\langle 1 2\rangle^2 \langle 2 3\rangle^2 \langle 3 1 \rangle^2\notag\\
    &\langle T^{-}T^{-}T^{+}\rangle=c_T\bigg(\frac{(E-2p_3)^2(E^3-E(p_1p_2+p_2p_3+p_3p_1)-p_1p_2p_3}{E^2p_1^2p_2^2p_3^2}\bigg)\langle 1 2\rangle^2\langle 2 \Bar{3}\rangle^2\langle \Bar{3}1\rangle^2.\notag
\end{align}
\end{widetext}
$c_E$ and $c_O$ are the partiy even and parity odd homogeneous coefficients whereas $c_T$ is the parity even nonhomogeneous coefficient. The remaining helicities can be obtained via exchanges or by the complex conjugation. 

Just like the previous example of $\langle JJJ\rangle$, while the homogeneous contribution is in accordance with statement $1$ \eqref{statement1}, the nonhomogeneous part seems to violate statement $2$ \eqref{statement2} as it is nonzero in the $(---)$ and $(+++)$ helicities. However, we shall show that by adding a suitable contact term, we can render such contributions zero. Writing an ansatz for the contact term and fixing the coefficients by demanding that it should cancel out the nonhomogeneous contribution in the $(---)$ and $(+++)$ helicities we find,
\begin{widetext}
\small
 \begin{align}\label{TTTcontact}
    &\langle TTT\rangle_{\text{contact}}=c_T\bigg[\frac{8(z_1\cdot p_2)(z_1\cdot z_2)(z_2\cdot z_3)(z_3\cdot p_1)(E+p_2)}{3}-\frac{8(z_1\cdot p_2)(z_1\cdot z_3)(z_2\cdot p_1)(z_2\cdot z_3)}{3}(E+p_3)\notag\\
    &-\frac{8(z_1\cdot z_2)(z_1\cdot z_3)(z_2\cdot p_1)(z_3\cdot p_1)}{3}(E+p_1)-8(z_1\cdot z_2)(z_1\cdot z_3)(z_2\cdot z_3)\bigg(p_1^3+2p_1^2(p_2+p_3)+2p_1(p_2^2+p_3^2)+(p_2+p_3)(p_2^2+p_2p_3+p_3^2)\bigg)\bigg].
\end{align}
\normalsize
   
\end{widetext}

By redefining,
\begin{align}
    \langle TTT\rangle=\langle TTT\rangle-\langle TTT\rangle_{\text{contact}},
\end{align}
we find,
\small
    \begin{align}
    &\langle T^{-}T^{-}T^{-}\rangle=\bigg(\frac{(c_E+ic_O) p_1 p_2 p_3}{E^6}\bigg)\langle 1 2\rangle^2 \langle 2 3\rangle^2 \langle 3 1 \rangle^2,\notag\\
    &\langle T^{-}T^{-}T^{+}\rangle=c_T\frac{-2}{3p_1^2p_2^2p_3E^2}\bigg(5p_1^4+3p_1^3(4p_2+3p_3)\notag\\&+p_1^2(14p_2^2+11p_2p_3+9p_3^2)+p_2(5p_2^3+9p_2^2p_3+9 p_2p_3^2+5p_3^3)\notag\\&+p_1(12p_2^3+11p_2^2p_3+16 p_2p_3^2+5 p_3^3\bigg)\langle 1 2\rangle^2\langle 2 \Bar{3}\rangle^2\langle \Bar{3} 1\rangle^2,\notag
\end{align}
\normalsize
proving that the contact term \eqref{TTTcontact}, renders the nonhomogeneous $\langle TTT\rangle$ zero in the $(- - -)$ and $(+++)$ helicities.

We now turn to an example where there also exists a zero-helicity configuration.
\subsection*{$\textbf{Example~3}:\langle TJJ\rangle$}
The helicity decomposition of this correlator is as follows \footnote{\label{ft:A}This expression can easily be obtained from the free theory computations using the definitions provided in \eqref{handnh}.}:
\small
\begin{align}\label{TJJmmmandpmm}
    &\langle T^{-}J^{-}J^{-}\rangle=\frac{(c_1+ic_1')p_1}{E^4}\langle 1 2\rangle^2\langle 3 1 \rangle^2-c_J\frac{\langle 1 2\rangle^2\langle 3 1\rangle^2(p_2+p_3)}{p_1^2 p_2 p_3},\notag\\
    &\langle T^{+}J^{-}J^{-}\rangle=-c_J\frac{\langle 2 3\rangle^4(p_1^2-(p_2-p_3)^2)^2(p_2+p_3)}{\langle 1 2\rangle^2\langle 3 1\rangle^2 p_1^2 p_2 p_3}\\
    &\langle T^{-}J^{-}J^{+}\rangle=-
    c_J\frac{\langle 1 2\rangle^4}{3E^2\langle 2 3\rangle^2 p_1^2 p_2 p_3}(E-2p_1)^2\big(3E^2(E-p_1)\notag\\&~~~~~~~~~~~~~~~~+8(p_1^2-E^2)p_2+8(p_1+E)p_2^2\big).\notag
\end{align}
\normalsize
$c_1$ and $c_1'$ are the parity even and odd homogeneous coefficients while $c_J$ is the coefficient of the nonhomogeneous parity even contribution.

It seems that not only does $\langle TJJ\rangle$ contain a nonhomogeneous contribution in the $(---)$ and $(+++)$ helicity configurations but also a zero helicity configuration, both of which should not exist according to statements $2$ \eqref{statement2} and $3$ \eqref{statement3}. 

By writing an ansatz for contact terms to cancel these contributions, we were able to find one that simultaneously cancels out both of them:
\begin{align}\label{TJJcontact}
\langle TJJ\rangle_{\text{contact}}=-4c_J(z_1\cdot z_2)(z_1\cdot z_3)(p_2+p_3).
\end{align}
By redefining,
\begin{align}
    \langle TJJ\rangle_{new}=\langle TJJ\rangle-\langle TJJ\rangle_{\text{contact}}\,,
\end{align}
we find
\begin{align}\label{TJJnhmmmpmm}
    &\langle T^{-}J^{-}J^{-}\rangle_{new}=\frac{(c_1+ic_1')p_1}{E^4}\langle 1 2\rangle^2\langle 3 1 \rangle^2\notag\\
    &\langle T^{+}J^{-}J^{-}\rangle_{new}=0\notag\\
    &\langle T^{-}J^{-}J^{+}\rangle_{new}=\frac{8c_J\langle 1 2\rangle^4}{3\langle 2 3\rangle^2 p_1^2 E^2}(E-2p_1)^2(E+p_1)\,.
\end{align}
Thus, the simple contact term \eqref{TJJcontact} suffices to render the-nonhomogeneous $\langle TJJ\rangle$ zero in the $(- - -)$ and $(+++)$ helicities as well as the zero-helicity configurations.

\subsection*{$\textbf{Example~4}:\langle J_3 T J\rangle$}
Let us now consider a correlator involving a higher-spin current to show that our statements also hold in such cases as well. Since the homogeneous part of this correlator obeys statement $1$ \eqref{statement1}, we focus on the apparent discrepancy between this correlator's nonhomogeneous part with our statements $2$ \eqref{statement2} and $3$ \eqref{statement3}. The nonhomogeneous part of this correlator in the $(- - -)$ and $(+ - -)$ configurations are respectively given by\footnote{Again, one can compute this via Wick contractions and use the definition \eqref{handnh}},
\begin{align}
    &\langle J_3^{-}T^{-}J^{-}\rangle_{nh}=a\frac{-\langle 1 2\rangle^4\langle 3 1\rangle^2}{p_1^3 p_2^2 p_3}\big( 6p_2^3-p_1^2p_3\notag\\&+4p_2(p_1+2p_2)p_3-7(p_1-2p_2)p_3^2+14 p_3^3\big),\notag\\
    &\langle J_3^{+}T^{-}J^{-}\rangle_{nh}=a\frac{\langle 2 3\rangle^6(p_1+p_2-p_3)^2(p_1-p_2+p_3)^4}{\langle 1 2\rangle^2\langle 3 1\rangle^4 p_1^3 p_2^3 p_3}\notag\\&\big(-6 p_2^3-14p_3^3+(p_1^2+4p_1p_2-8p_2^2)p_3
    -7(p_1+2p_2)p_3^2\big)\,, \notag
\end{align}
$a$ is the coefficient of the parity even nonhomogeneous contribution. 
Although these expressions are quite complicated, we were able to write an ansatz and determine a contact term that can simultaneously cancel both these expressions, rendering the nonhomogeneous contribution zero in the $(---)$, $(+++)$ helicities as well as the two zero-helicity configurations. The explicit expression for the contact term is as follows:
\begin{align*}
    &\langle J_3 T J\rangle_{\text{contact}}=4a(z_1\cdot z_2)\bigg(4(z_1\cdot p_2)\big(9(z_2\cdot p_1)(z_1\cdot z_3)\notag\\&-7(z_1\cdot p_2)(z_2\cdot z_3)\big)p_3+(z_1\cdot z_2)\big(-14(z_3\cdot p_1)(z_1\cdot p_2)p_3\notag\\&+(z_1\cdot z_3)(12p_2^3+9p_1^2p_3+13 p_2^2 p_3+3 p_3^3)\big)\bigg).
\end{align*}
By redefining,
\begin{align}
    \langle J_3 T J\rangle_{nh,new}=\langle J_3 T J\rangle_{nh}-\langle J_3 T J\rangle_{\text{contact}},
\end{align}
we find,
\begin{align}
    \langle J_3^{-}T^{-}J^{-}\rangle_{nh,new}=\langle J_3^{+}T^{-}J^{-}\rangle_{nh,new}=0.
\end{align}
With this, we conclude our examples and attempt to generalize our findings to arbitrary correlators inside the triangle.

\subsection*{Conclusion}
In all the examples that we considered, we showed that it is always possible to add contact terms to the correlators to render the nonhomogeneous contributions zero in the $(- - -)$, $(+ + +)$ helicities as well as the zero-helicity configuration where applicable. Essentially, what this means is that the Ward-Takahashi identity for the correlators in these helicity sectors can be removed via re-defining the correlator by adding particular contact terms. If we take this as an input, i.e. the Ward-Takahashi identities being trivial in these helicity configurations, one can prove in general that the nonhomogeneous contribution to the correlator vanishes in the $(- - -)$, $(+ + +)$ and zero-helicity sector. This follows from the statement that there exists a unique solution (that is consistent with the operator product expansion \cite{Jain:2021whr}) to the conformal Ward identities in these configurations, which are the homogeneous correlators. We also investigated correlators involving half-integer spin currents in appendix \ref{appendix:halfintcase}, where we find a perfect agreement with our statements. Given these facts, it would be nice to rigorously prove statements 2 \eqref{statement2} and 3 \eqref{statement3} but we defer such an analysis to the future.

Thus, after all the dust has settled, we see that using statements $1$, $2$ and $3$, \eqref{statement1}, \eqref{statement2} and \eqref{statement3}), the expressions for three-point functions inside the triangle for generic CFTs in the eight helicity configurations are,
\begin{widetext}
\begin{align}\label{insidetriangles1s2s3}
    &\langle J_{s_1}^{-}J_{s_2}^{-}J_{s_3}^{-}\rangle=(n_f-n_b-2in_{\text{odd}})\langle J_{s_1}^{-}J_{s_2}^{-}J_{s_3}^{-}\rangle_h,~\,\langle J_{s_1}^{+}J_{s_2}^{+}J_{s_3}^{+}\rangle=(n_f-n_b+2in_{\text{odd}})\langle J_{s_1}^{+}J_{s_2}^{+}J_{s_3}^{+}\rangle_h,\notag\\
    &\langle J_{s_1}^{-}J_{s_2}^{-}J_{s_3}^{+}\rangle=(n_b+n_f)\langle J_{s_1}^{-}J_{s_2}^{-}J_{s_3}^{+}\rangle_{nh},~~~~~~~~~~~~\langle J_{s_1}^{+}J_{s_2}^{+}J_{s_3}^{-}\rangle=(n_b+n_f)\langle J_{s_1}^{+}J_{s_2}^{+}J_{s_3}^{-}\rangle_{nh},\notag\\
    &\langle J_{s_1}^{-}J_{s_2}^{+}J_{s_3}^{-}\rangle=(n_b+n_f)\langle J_{s_1}^{-}J_{s_2}^{+}J_{s_3}^{-}\rangle_{nh},~~~~~~~~~~~~\langle J_{s_1}^{+}J_{s_2}^{-}J_{s_3}^{+}\rangle=(n_b+n_f)\langle J_{s_1}^{+}J_{s_2}^{-}J_{s_3}^{+}\rangle_{nh},\notag\\
    &\langle J_{s_1}^{+}J_{s_2}^{-}J_{s_3}^{-}\rangle=(n_b+n_f)\langle J_{s_1}^{+}J_{s_2}^{-}J_{s_3}^{-}\rangle_{nh},~~~~~~~~~~~~\langle J_{s_1}^{-}J_{s_2}^{+}J_{s_3}^{+}\rangle=(n_b+n_f)\langle J_{s_1}^{-}J_{s_2}^{+}J_{s_3}^{+}\rangle_{nh},
\end{align}
\end{widetext}
and if there are any configurations with net zero helicity, the correlators are identically zero. We now turn to the study of three-point functions that are outside the triangle.
\subsubsection{\textbf{Three-point functions that are outside the triangle}}
Three-point functions of conserved currents that violate the spin triangle inequality take the following form \cite{Giombi:2011rz}:
\begin{align}
    \langle J_{s_1}J_{s_2}J_{s_3}\rangle=n_f \langle J_{s_1}J_{s_2}J_{s_3}\rangle_{FF}+n_b \langle J_{s_1}J_{s_2}J_{s_3}\rangle_{FB}\,,
\end{align}
i.e. in contrast to \eqref{3ptCFT}, the parity odd structure is incompatible with current conservation.\footnote{There can exist parity odd nonhomogeneous contributions but these are all contact term contributions. For instance, take the FF or FB Ward-Takahashi identity and replace all the two-point functions appearing there by their parity odd counterparts. This gives rise to two new structures that source these parity odd Ward-Takahashi identities. It would be interesting to see the consequences of the existence of these structures but we leave such an exercise for the future.} It is also interesting to note that in contrast to the situation inside the triangle \eqref{3ptCFThandnh}, there is no homogeneous correlator outside the triangle. The reason for this is that the free bosonic and free fermionic Ward-Takahashi identities are different outside the triangle \cite{Jain:2021whr}. It is also important to note that the only non-zero Ward-Takahashi identity is only due to the highest spin current in the correlator. For example, consider a correlator with a spin four current and two spin one currents, i.e., $\langle J_4 J_1 J_1\rangle$. Here, the only Ward-Takahashi identity is with respect to the spin-four current.\\

Just as we have made for three-point functions inside the triangle, we shall make some statements for their counterparts that are outside the triangle.
Without loss of generality, consider a correlator $\langle J_{s_1}J_{s_2}J_{s_3}\rangle$ with $s_1>s_2+s_3$, $s_2\ge s_3$. We have then, the following statements for correlators that violate the spin triangle inequality:\\

\textbf{\underline{Statement 4:}} The $FF-FB$ correlators have support iff $J_{s_2}$ and $J_{s_3}$ have the same helicities.
\begin{align}\label{statement4}
    &\langle J_{s_1}^{h_1}J_{s_2}^{-}J_{s_3}^{-}\rangle_{FF-FB}\ne 0\,, \qquad  \langle J_{s_1}^{h_1}J_{s_2}^{+}J_{s_3}^{+}\rangle_{FF-FB}\ne 0\notag\\
    &\langle J_{s_1}^{h_1}J_{s_2}^{-}J_{s_3}^{+}\rangle_{FF-FB}=\langle J_{s_1}^{h_1}J_{s_2}^{+}J_{s_3}^{-}\rangle_{FF-FB}=0.
\end{align}

\textbf{\underline{Statement 5:}} The $FF+FB$ correlators identically vanish when the helicities of $J_{s_2}$ and $J_{s_3}$ coincide.
\begin{align}\label{statement5}
    &\langle J_{s_1}^{h_1}J_{s_2}^{-}J_{s_3}^{-}\rangle_{FF+FB}= 0\,,\qquad  \langle J_{s_1}^{h_1}J_{s_2}^{+}J_{s_3}^{+}\rangle_{FF+FB}= 0\notag\\
    &\langle J_{s_1}^{h_1}J_{s_2}^{-}J_{s_3}^{+}\rangle_{FF+FB}\ne 0\,, \qquad \langle J_{s_1}^{h_1}J_{s_2}^{+}J_{s_3}^{-}\rangle_{FF+FB}\ne 0\,.
\end{align}

We verify statements $4$ and $5$ in appendix \ref{appendix:outsidetriangleexamples} via a few examples. It would be nice to find a general proof of these statements in the future. For statement $4$, however, we present a different argument for its validity later following as an implication of the next statement to be discussed below.\\

If we allow the current conservation to be ``slightly'' broken we instead have \cite{Giombi:2011rz},
\small
\begin{align}\label{s1s2s3outsidetriangle}
    \langle J_{s_1}J_{s_2}J_{s_3}\rangle&=n_f \langle J_{s_1}J_{s_2}J_{s_3}\rangle_{FF}+n_{odd}\langle J_{s_1}J_{s_2}J_{s_3}\rangle_{odd}\notag\\&+n_b \langle J_{s_1}J_{s_2}J_{s_3}\rangle_{FB},
\end{align}
\normalsize
where the odd piece is the one that is not conserved but rather, slightly-broken.
However, it was shown in \cite{Jain:2021gwa} that the odd part can be given by an epsilon transform of the difference of the fermionic and bosonic correlators:
\begin{align}\label{s1s2s3outsidetrianglewithEPT}
    \langle J_{s_1}J_{s_2}J_{s_3}\rangle&=n_f \langle J_{s_1}J_{s_2}J_{s_3}\rangle_{FF}+n_{odd}~\epsilon\cdot \langle J_{s_1}J_{s_2}J_{s_3}\rangle_{FF-FB}\notag\\&+n_b \langle J_{s_1}J_{s_2}J_{s_3}\rangle_{FB}\,.
\end{align}
The question now is, epsilon transform with respect to what? The answer is provided in the below statement:\\

\textbf{\underline{Statement 6:}} The epsilon transform required to obtain the odd correlator can be performed only with respect to the lowest two spins in the correlator. The proof of statement $6$ can be found in appendix \ref{appendix:epToutsidetriangle}. For example consider $\langle J_{s_1}J_{s_2}J_{s_3}\rangle$ with $s_1>s_2$, $s_2\ge s_3$. Then,
\small
\begin{align}\label{statement6}
    \langle J_{s_1}J_{s_2}J_{s_3}\rangle_{\text{odd}}=\langle J_{s_1}\epsilon\cdot J_{s_2}J_{s_3}\rangle_{\text{FF-FB}}=\langle J_{s_1}J_{s_2}\epsilon\cdot J_{s_3}\rangle_{\text{FF-FB}}\,.
\end{align}
\normalsize

Let us now summarize the structure of three-point correlators outside the triangle while employing \eqref{s1s2s3outsidetrianglewithEPT} together with statements $4,5$ and $6$, \eqref{statement4}, \eqref{statement5} and \eqref{statement6} in mind.
\subsection*{Helicity decomposition of correlators outside the triangle}
 We see that the expressions for three-point functions outside the triangle for CFTs with the slightly-broken higher-spin symmetry in the eight helicity configurations are (we assume $s_1>s_2$, $s_2\ge s_3$ in the expression below)
 \begin{widetext}
\begin{align}\label{outsidetriangles1s2s3}
    &\langle J_{s_1}^{-}J_{s_2}^{-}J_{s_3}^{-}\rangle=\tfrac{n_f-n_b-2i n_{odd}}{2}\langle J_{s_1}^{-}J_{s_2}^{-}J_{s_3}^{-}\rangle_{FF-FB}, &&\langle J_{s_1}^{+}J_{s_2}^{+}J_{s_3}^{+}\rangle=\tfrac{n_f-n_b+2i n_{odd}}{2}\langle J_{s_1}^{+}J_{s_2}^{+}J_{s_3}^{+}\rangle_{FF-FB},\notag\\
    &\langle J_{s_1}^{-}J_{s_2}^{-}J_{s_3}^{+}\rangle=\tfrac{n_f+n_b}{2}\langle J_{s_1}^{-}J_{s_2}^{-}J_{s_3}^{+}\rangle_{FF+FB},&&\langle J_{s_1}^{+}J_{s_2}^{+}J_{s_3}^{-}\rangle=\tfrac{n_f+n_b}{2}\langle J_{s_1}^{+}J_{s_2}^{+}J_{s_3}^{-}\rangle_{FF+FB},\notag\\
    &\langle J_{s_1}^{-}J_{s_2}^{+}J_{s_3}^{-}\rangle=\tfrac{n_f+n_b}{2}\langle J_{s_1}^{-}J_{s_2}^{+}J_{s_3}^{-}\rangle_{FF+FB},&&\langle J_{s_1}^{+}J_{s_2}^{-}J_{s_3}^{+}\rangle=\tfrac{n_f+n_b}{2}\langle J_{s_1}^{+}J_{s_2}^{-}J_{s_3}^{+}\rangle_{FF+FB},\notag\\
    &\langle J_{s_1}^{+}J_{s_2}^{-}J_{s_3}^{-}\rangle=\tfrac{n_f-n_b-2i n_{odd}}{2}\langle J_{s_1}^{+}J_{s_2}^{-}J_{s_3}^{-}\rangle_{FF-FB},&&\langle J_{s_1}^{-}J_{s_2}^{+}J_{s_3}^{+}\rangle=\tfrac{n_f-n_b+2i n_{odd}}{2}\langle J_{s_1}^{-}J_{s_2}^{+}J_{s_3}^{+}\rangle_{FF-FB}.
\end{align}
\end{widetext}

Let us now turn to an explicit example of a CFT with the slightly-broken higher-spin symmetry, i.e. to Chern-Simons matter theories.

\subsection{Quasi-Fermionic Theory}\label{sec:CS+fermions}
The theory of interest is $SU(N_f)$ Chern-Simons+fundamental fermions (henceforth called the quasi-fermionic (QF) theory).  The action of the QF theory is,
\small
\begin{align}
    S_{FF,CS}=\int d^3 x\bigg(\bar{\psi}\slashed{D}\psi+i\epsilon^{\mu\nu\rho}\frac{\kappa_f}{4\pi}\Tr{A_\mu\partial_\nu A_\rho-\frac{2i}{3}A_\mu A_\nu A_\rho}\bigg).
\end{align}
\normalsize
As usual, we take the large $N_f\to\infty $ limit. We also take the limit $\kappa_f\to \infty$ keeping the t'Hooft coupling $\lambda_f=\frac{N_f}{\kappa_f}$ fixed. The spectrum of this theory includes a parity odd $\Delta=2+\order{\frac{1}{N_f}}$ scalar $O_2$, conserved spin-$1$ and spin-$2$ currents $J$ and $T$ and weakly nonconserved currents $J_s,s=3,4,\cdots$ with $\Delta_s=s+1+\order{\frac{1}{N_f}}$.

This theory enjoys a strong-weak duality with the  $SU(N_b)$ critical bosonic (CB) theory, which is the theory obtained by flowing to the Wilson-Fisher fixed point starting with the following action:
\begin{align}
    &S_{CB,CS}=\int d^3 x\bigg(D_\mu \bar{\phi}D_\mu \phi+\frac{\lambda_4}{N_b}(\bar{\phi}\phi)^2\notag\\&+i\epsilon^{\mu\nu\rho}\frac{\kappa_b}{4\pi}\Tr{A_\mu\partial_\nu A_\rho-\frac{2i}{3}A_\mu A_\nu A_\rho}\bigg).
\end{align}
We take the limits $N_b\to\infty,\lambda_4\to \infty$ and $\kappa_b\to \infty$ keeping fixed $\frac{\lambda_4}{N_b}$ and $\lambda_b=\frac{N_b}{\kappa_b}$.  The spectrum of this theory includes a parity even $\Delta=2+\order{\frac{1}{N_b}}$ scalar $O_2$, conserved spin-$1$ and spin-$2$ currents $J$ and $T$ and weakly nonconserved currents $J_s,s=3,4,\cdots$ with $\Delta=s+1+\order{\frac{1}{N_b}}$\footnote{A CB three point correlator with a $O_2$ insertion can be obtained via a Legendre transform of a FB correlator with an $O_1$ insertion instead of the $O_2$ operator. Three point correlators without scalar insertions coincide in the FB and CB theory and thus we will use the labels interchangeably in such cases.}. 

The nonconservation is not seen at the level of two-point correlation functions, which take the form of the exactly conserved case, \eqref{JsJs2point} with $c_{s,even}=\Tilde{N}$ and $c_{s,odd}=\Tilde{N}\Tilde{\lambda}$. Three-point functions inside and outside the triangle respectively take the form of \eqref{3ptCFT} and \eqref{s1s2s3outsidetriangle} with the coefficients are given by \cite{Maldacena:2012sf}\footnote{We define $\Tilde{N}$ and $\Tilde{\lambda}$ in \eqref{tildevariables}.}
\begin{align}\label{nfnoddnbcoeffs}
    n_f=\frac{\Tilde{N}}{1+\Tilde{\lambda}^2}\,,\qquad n_{odd}=\frac{\Tilde{N}\Tilde{\lambda}}{1+\Tilde{\lambda}^2}\,,\qquad n_b=\frac{\Tilde{N}\Tilde{\lambda}^2}{1+\Tilde{\lambda}^2}\,.
\end{align}
If the three-point function is inside the triangle, then the correlator is conserved with respect to all the currents. However, when the spin triangle inequality is violated, the odd part of the correlator gives rise to the nonconservation.\\

Thus, with the three-point correlators both inside and outside the triangle under our control, we use the results of this section in the quest to identify the (anti-)chiral limit in Chern-Simons matter theory.

\section{(Anti-)chiral limit for Chern-Simons matter theory}\label{sec:ChiralHSfromCS}
Our goal in this section is to find a closed subsector of Chern-Simons matter theories that is dual to (anti-)Chiral higher-spin gravity in $\text{AdS}_4$. In the main text, we focus on the QF theory. For the other slightly broken higher spin theory, that is the theory of Chern-Simons+fundamental bosons, please refer to appendix \ref{appendix:QBchirallimit}.

The limit that we take to reach the (anti-)Chiral theory should be a nonunitary one as the (anti-)chiral bulk theory is nonunitary. To find out the correct prescription, we begin by analyzing correlation functions of the single trace primary operators in the Chern-Simons matter theory.

Before we proceed, let us set the notation straight.
In contrast to \cite{Jain:2022ajd}, we normalize our single trace operators such that their two-point functions are $\order{1}$, with an additional rescaling as follows:
\begin{align}\label{rescaling}
    &\Tilde{J}_s^{\pm}=\frac{e^{\mp i \theta}}{\sqrt{\tilde{N}}}J_s^{\pm}\notag\\
    &\Tilde{O}_2=\frac{\cos{\theta}}{\sqrt{\tilde{N}}}~O_2,
\end{align}
where $\theta=\frac{\pi \lambda}{2}$ and $\lambda=\frac{N}{k}$ is the 't Hooft coupling in the Chern-Simons matter theory. Note that the rescaling of the currents is different for their positive and negative components. For real $\theta$ this is required as $J_s^{+}$ and $J_s^{-}$ are complex conjugates of each other. However, if we complexify $\theta$ this is no longer true and $J_s^{+}$ and $J_s^{-}$ become independent quantities. Finally, we define for convenience,
\begin{align}\label{tildevariables}
    &\Tilde{N}=2N\frac{\sin{2\theta}}{2\theta}\,, &
    &\Tilde{\lambda}=\tan{\theta}\,.
\end{align}
We now state two limits that, as we shall show, project us into the anti-chiral or chiral sectors.
\small
\begin{align}\label{anti-chirallimit}
    \textbf{anti-chiral limit:~} (\Tilde{N}\to \infty, \Tilde{\lambda}\to -i\equiv\theta\to -i\infty)~\textbf{with~}\frac{e^{i\theta}}{\sqrt{\Tilde{N}}}=g_{ac}, 
\end{align}
\normalsize
and,
\small
\begin{align}\label{chirallimit}
    \textbf{chiral limit:~} (\Tilde{N}\to \infty, \Tilde{\lambda}\to i\equiv\theta\to i\infty)~\textbf{with~}\frac{e^{-i\theta}}{\sqrt{\Tilde{N}}}=g_{c}.
\end{align}
\normalsize
Let us begin by investigating the consequences of these limits for two-point functions.
\subsection{Two-point functions}
The two-point functions of the spin-$s$ currents in the QF theory are given by,
\begin{align}
    \langle J_s J_s\rangle_{QF}=\langle J_s J_s\rangle_{even}+\Tilde{\lambda}\langle J_s J_s\rangle_{odd}\,,
\end{align}
where the odd piece is given in terms of the even one through an epsilon transformation as we discussed in section \ref{sec:CFT3corrfacts}.
In spinor-helicity variables, the correlator takes the following form in the two nonzero helicity configurations (note that this and all other expressions in spinor-helicity variables are after the rescaling \eqref{rescaling}):
\begin{align}\label{2pointspins}
    &\langle \Tilde{J}_s^{-}\Tilde{J}_s^{-}\rangle_{QF}=(1
    -i\Tilde{\lambda})\langle J_s^{-}J_s^{-}\rangle_{even}\notag\\&,\langle \Tilde{J}_s^{+}\Tilde{J}_s^{+}\rangle_{QF}=(1+i\Tilde{\lambda})\langle J_s^{+}J_s^{+}\rangle_{even}.
\end{align}
Let us now take the limit $\Tilde{\lambda}\to -i$ in \eqref{2pointspins}. In this limit, only the $(- -)$ configuration survives. We obtain,
\begin{align}
    &\lim_{\Tilde{\lambda}\to -i}\langle \Tilde{J}_s^{-}\Tilde{J}_s^{-}\rangle_{QF}=2\langle J_s^{-}J_s^{-}\rangle_{even}=2\frac{\langle 1 2\rangle^{2s}}{p}\,,\notag\\& \lim_{\Tilde{\lambda}\to -i}\langle\Tilde{J}_s^{+}\Tilde{J}_s^{+}\rangle_{QF}=0.
\end{align}
Similarly, as $\Tilde{\lambda}\to i$, we obtain only the $(+ +)$ helicity configuration. The story for the scalar two-point function is simple: it reads,
\begin{align}
    \langle \Tilde{O}_2 \Tilde{O}_2\rangle_{QF}=\langle O_2 O_2\rangle_{FF}\,.
\end{align}
Let us move on to the three-point case where we can test this limit further.
\subsection{Three-point functions}
We will show that \eqref{anti-chirallimit} and \eqref{chirallimit} do indeed produce (anti-)chiral three-point functions. Let us begin with correlators that are inside the triangle.
\subsection*{Spinning correlators inside the triangle}
\subsection*{\textbf{Example~}1: $\langle TJJ\rangle$}
This correlator is given by,
\begin{align}\label{TJJcorr}
    \langle TJJ\rangle_{QF}&=\frac{1}{\sqrt{\Tilde{N}}(1+\Tilde{\lambda}^2)}\big(\langle TJJ\rangle_{FF}+\Tilde{\lambda}~\epsilon\cdot\langle TJJ\rangle_{FF-FB}\notag\\&+\Tilde{\lambda}^2\langle TJJ\rangle_{FB}\big).
\end{align}
Using our results for correlators inside the triangle \eqref{insidetriangles1s2s3} with $n_f,n_b$ and $n_{\text{odd}}$ given in \eqref{nfnoddnbcoeffs},  we obtain the expression of this correlator in the eight helicity configurations:
\small
\begin{align}
    &\langle \Tilde{T}^{-}\Tilde{J}^{-}\Tilde{J}^{-}\rangle=\frac{e^{i\theta}}{\sqrt{\Tilde{N}}}\langle T^{-}J^{-}J^{-}\rangle_h,\langle \Tilde{T}^{+}\Tilde{J}^{+}\Tilde{J}^{+}\rangle=\frac{e^{-i\theta}}{\sqrt{\Tilde{N}}}\langle T^{+}J^{+}J^{+}\rangle_h\notag\\
    &\langle \Tilde{T}^{-}\Tilde{J}^{-}\Tilde{J}^{+}\rangle=\frac{e^{i\theta}}{\sqrt{\Tilde{N}}}\langle T^{-}J^{-}J^{+}\rangle_{nh},\langle \Tilde{T}^{+}\Tilde{J}^{+}\Tilde{J}^{-}\rangle=\frac{e^{-i\theta}}{\sqrt{\Tilde{N}}}\langle T^{+}J^{+}J^{-}\rangle_{nh} \notag\\
    &\langle \Tilde{T}^{-}\Tilde{J}^{+}\Tilde{J}^{-}\rangle=\frac{e^{i\theta}}{\sqrt{\Tilde{N}}}\langle T^{-}J^{+}J^{-}\rangle_{nh},\langle \Tilde{T}^{+}\Tilde{J}^{-}\Tilde{J}^{+}\rangle=\frac{e^{-i\theta}}{\sqrt{\Tilde{N}}}\langle T^{+}J^{-}J^{+}\rangle_{nh}\notag\\
    &\langle \Tilde{T}^{-}\Tilde{J}^{+}\Tilde{J}^{+}\rangle=0~~~~~~~~~~~~~~~~~~~~~~~,\langle \Tilde{T}^{+}\Tilde{J}^{-}\Tilde{J}^{-}\rangle=0.
\end{align}
\normalsize
Let us now take the anti-chiral limit defined in \eqref{anti-chirallimit}. In this limit we obtain,
\begin{align}\label{TJJantichiral}
    &\langle \Tilde{T}^{-}\Tilde{J}^{-}\Tilde{J}^{-}\rangle=g_{ac}\langle T^{-}J^{-}J^{-}\rangle_h\,, \qquad&\langle \Tilde{T}^{+}\Tilde{J}^{+}\Tilde{J}^{+}\rangle&=0\notag\,, \\
    &\langle \Tilde{T}^{-}\Tilde{J}^{-}\Tilde{J}^{+}\rangle=g_{ac}\langle T^{-}J^{-}J^{+}\rangle_{nh}\,, \qquad&\langle \Tilde{T}^{+}\Tilde{J}^{+}\Tilde{J}^{-}\rangle&=0\notag\,, \\
    &\langle \Tilde{T}^{-}\Tilde{J}^{+}\Tilde{J}^{-}\rangle=g_{ac}\langle T^{-}J^{+}J^{-}\rangle_{nh}\,, \qquad&\langle \Tilde{T}^{+}\Tilde{J}^{-}\Tilde{J}^{+}\rangle&=0\notag\,, \\
    &\langle \Tilde{T}^{-}\Tilde{J}^{+}\Tilde{J}^{+}\rangle=0\,, &\langle \Tilde{T}^{+}\Tilde{J}^{-}\Tilde{J}^{-}\rangle&=0.
\end{align}
Equation \eqref{TJJantichiral} contains only the net negative helicity configurations of the correlators and is precisely the definition of the anti-chiral sector, showing that the limit \eqref{anti-chirallimit} is as we desired.
One point to note is that if we had not re-scaled the currents as in \eqref{rescaling}, we would only have been able to retain the $(- - -)$ helicity configuration. However, with the rescaling, we are able to obtain the full anti-chiral sector viz \eqref{TJJantichiral}. Let us consider another example now to further illustrate and validate \eqref{anti-chirallimit}:
\subsection*{\textbf{Example~}2: $\langle TTT\rangle$}
This correlator is given by
\footnote{There can, in principle, exist a parity odd nonhomogeneous contact term. If we allow for such contributions, $\langle TTT\rangle_{nh,\text{even}}$ can be cancelled out by $\langle TTT\rangle_{nh,\text{odd}}$ in the $(---)$ and $(+++)$ helicities as can be seen given the expressions in \cite{Jain:2021vrv}. However, for correlators such as $\langle TJJ\rangle$, there is no nonhomogeneous parity odd contact term possible and thus our parity even contact term analysis of section \ref{sec:CFT3corrfacts} is essential.}
\small
\begin{align}\label{TTT}
    \langle TTT\rangle=\frac{1}{\sqrt{\Tilde{N}}(1+\Tilde{\lambda}^2)}\big(\langle TTT\rangle_{FF}+\epsilon\cdot\langle TTT\rangle_{FF-FB}+\Tilde{\lambda}^2\langle TTT\rangle_{FB}\big).
\end{align}
\normalsize
Using \eqref{insidetriangles1s2s3} and \eqref{nfnoddnbcoeffs}, we see that this correlator in the eight helicity configurations reads
\footnotesize
\begin{align}
    &\langle \Tilde{T}^{-}\Tilde{T}^{-}\Tilde{T}^{-}\rangle=\frac{e^{i\theta}}{\sqrt{\Tilde{N}}}\langle T^{-}T^{-}T^{-}\rangle_h,\langle \Tilde{T}^{+}\Tilde{T}^{+}\Tilde{T}^{+}\rangle=\frac{e^{-i\theta}}{\sqrt{\Tilde{N}}}\langle T^{+}T^{+}T^{+}\rangle_h\notag\\
    &\langle \Tilde{T}^{-}\Tilde{T}^{-}\Tilde{T}^{+}\rangle=\frac{e^{i\theta}}{\sqrt{\Tilde{N}}}\langle T^{-}T^{-}T^{+}\rangle_{nh},\langle \Tilde{T}^{+}\Tilde{T}^{+}\Tilde{T}^{-}\rangle=\frac{e^{-i\theta}}{\sqrt{\Tilde{N}}}\langle T^{+}T^{+}T^{-}\rangle_{nh}\notag\\
    &\langle \Tilde{T}^{-}\Tilde{T}^{+}\Tilde{T}^{-}\rangle=\frac{e^{i\theta}}{\sqrt{\Tilde{N}}}\langle T^{-}T^{+}T^{-}\rangle_{nh},\langle \Tilde{T}^{+}\Tilde{T}^{-}\Tilde{T}^{+}\rangle=\frac{e^{-i\theta}}{\sqrt{\Tilde{N}}}\langle T^{+}T^{-}T^{+}\rangle_{nh}\notag\\
    &\langle \Tilde{T}^{+}\Tilde{T}^{-}\Tilde{T}^{-}\rangle=\frac{e^{i\theta}}{\sqrt{\Tilde{N}}}\langle T^{+}T^{-}T^{-}\rangle_{nh},\langle \Tilde{T}^{-}\Tilde{T}^{+}\Tilde{T}^{+}\rangle=\frac{e^{-i\theta}}{\sqrt{\Tilde{N}}}\langle T^{-}T^{+}T^{+}\rangle_{nh}.
\end{align}
\normalsize
Taking the anti-chiral limit \eqref{anti-chirallimit} we obtain only the anti-chiral helicity configurations as desired:
\begin{align}\label{TTTantichiral}
     &\langle \Tilde{T}^{-}\Tilde{T}^{-}\Tilde{T}^{-}\rangle=g_{ac}\langle T^{-}T^{-}T^{-}\rangle_h\,, &&\langle \Tilde{T}^{+}\Tilde{T}^{+}\Tilde{T}^{+}\rangle=0\notag\\
    &\langle \Tilde{T}^{-}\Tilde{T}^{-}\Tilde{T}^{+}\rangle=g_{ac}\langle T^{-}T^{-}T^{+}\rangle_{nh}\,, &&\langle \Tilde{T}^{+}\Tilde{T}^{+}\Tilde{T}^{-}\rangle=0\notag\\
    &\langle \Tilde{T}^{-}\Tilde{T}^{+}\Tilde{T}^{-}\rangle=g_{ac}\langle T^{-}T^{+}T^{-}\rangle_{nh}\,, &&\langle \Tilde{T}^{+}\Tilde{T}^{-}\Tilde{T}^{+}\rangle=0\notag\\
    &\langle \Tilde{T}^{+}\Tilde{T}^{-}\Tilde{T}^{-}\rangle=g_{ac}\langle T^{+}T^{-}T^{-}\rangle_{nh}\,, &&\langle \Tilde{T}^{-}\Tilde{T}^{+}\Tilde{T}^{+}\rangle=0.
\end{align}
We note that if the nonhomogeneous correlator was nonzero in the $(- - -)$ helicity configuration, we would have a contribution of the form $\frac{e^{3i\theta}}{\sqrt{\Tilde{N}}}\langle T^{-}T^{-}T^{-}\rangle_{nh}$ which would be exponentially enhanced compared to all the other helicity configurations in the anti-chiral limit. It being zero via the addition of contact terms like \eqref{TTTcontact} is essential to obtain a uniform scaling for all the anti-chiral helicity configurations as in \eqref{TTTantichiral}.

Let us now consider a correlator outside the triangle and investigate if our findings carry over.
\subsection*{Spinning correlators outside the triangle}
\subsection*{\textbf{Example~}3: $\langle J_4 J_1 J_1\rangle$}
Using \eqref{outsidetriangles1s2s3} and \eqref{nfnoddnbcoeffs}, we see that the eight helicity configurations of this correlator are,
\begin{widetext}
\begin{align}
    &\langle \Tilde{J}_4^{-}\Tilde{J}^{-}\Tilde{J}^{-}\rangle=\frac{e^{i\theta}}{2\sqrt{\Tilde{N}}}\langle J_4^{-}J^{-}J^{-}\rangle_{FF-FB}\,,&&\langle \Tilde{J}_4^{+}\Tilde{J}^{+}\Tilde{J}^{+}\rangle=\frac{e^{-i\theta}}{2\sqrt{\Tilde{N}}}\langle J_4^{+}J^{+}J^{+}\rangle_{FF-FB}\notag\\
    &\langle \Tilde{J}_4^{-}\Tilde{J}^{-}\Tilde{J}^{+}\rangle=\frac{e^{i\theta}}{2\sqrt{\Tilde{N}}}\langle J_4^{-}J^{-}J^{+}\rangle_{FF+FB}\,,&&\langle \Tilde{J}_4^{+}\Tilde{J}^{+}\Tilde{J}^{-}\rangle=\frac{e^{-i\theta}}{2\sqrt{\Tilde{N}}}\langle J_4^{+}J^{+}J^{-}\rangle_{FF+FB}\notag\\
    &\langle \Tilde{J}_4^{-}\Tilde{J}^{+}\Tilde{J}^{-}\rangle=\frac{e^{i\theta}}{2\sqrt{\Tilde{N}}}\langle J_4^{-}J^{+}J^{-}\rangle_{FF+FB}\,,&&\langle \Tilde{J}_4^{+}\Tilde{J}^{-}\Tilde{J}^{+}\rangle=\frac{e^{-i\theta}}{2\sqrt{\Tilde{N}}}\langle J_4^{+}J^{-}J^{+}\rangle_{FF+FB}\notag\\
    &\langle \Tilde{J}_4^{-}\Tilde{J}^{+}\Tilde{J}^{+}\rangle=\frac{e^{i\theta}}{2\sqrt{\Tilde{N}}}\langle J_4^{-}J^{+}J^{+}\rangle_{FF-FB}\,,&&\langle \Tilde{J}_4^{+}\Tilde{J}^{-}\Tilde{J}^{-}\rangle=\frac{e^{-i\theta}}{2\sqrt{\Tilde{N}}}\langle J_4^{+}J^{-}J^{-}\rangle_{FF-FB}\,.
\end{align}
\end{widetext}
Let us now take the anti-chiral limit defined in \eqref{anti-chirallimit}. In this limit we obtain,
\begin{align}
    &\langle \Tilde{J}_4^{-}\Tilde{J}^{-}\Tilde{J}^{-}\rangle=\frac{g_{ac}}{2}\langle J_4^{-}J^{-}J^{-}\rangle_{FF-FB}\,,&&\langle \Tilde{J}_4^{+}\Tilde{J}^{+}\Tilde{J}^{+}\rangle=0\notag\\
    &\langle \Tilde{J}_4^{-}\Tilde{J}^{-}\Tilde{J}^{+}\rangle=\frac{g_{ac}}{2}\langle J_4^{-}J^{-}J^{+}\rangle_{FF+FB}\,,&&\langle \Tilde{J}_4^{+}\Tilde{J}^{+}\Tilde{J}^{-}\rangle=0\notag\\
    &\langle \Tilde{J}_4^{-}\Tilde{J}^{+}\Tilde{J}^{-}\rangle=\frac{g_{ac}}{2}\langle J_4^{-}J^{+}J^{-}\rangle_{FF+FB}\,,&&\langle \Tilde{J}_4^{+}\Tilde{J}^{-}\Tilde{J}^{+}\rangle=0\notag\\
    &\langle \Tilde{J}_4^{-}\Tilde{J}^{+}\Tilde{J}^{+}\rangle=\frac{g_{ac}}{2}\langle J_4^{-}J^{+}J^{+}\rangle_{FF-FB}\,,&&\langle \Tilde{J}_4^{+}\Tilde{J}^{-}\Tilde{J}^{-}\rangle=0\,,
\end{align}
which indeed is the anti-chiral sector. Note that we used the facts that the $FF+FB$ contribution to this correlator vanishes in the $(- - -)$ as well as the $(- + +)$ helicity configurations (as well as their complex conjugate helicities). We also used the fact that the $FF-FB$ contribution vanishes in the $(- - +)$ helicity as well as its complex conjugate and their $(2\leftrightarrow 3)$ permutations. These facts were explicitly obtained in \cite{Jain:2021whr}. Some comments about parity odd contact terms are also in order. The FF and FB Ward-Takahashi identities are distinct outside the triangle. Thus, if we take the $\langle JJ\rangle$ two-point functions appearing on the RHS of these Ward-Takahashi identities and replace them with the parity odd two-point function, we will obtain two distinct parity odd correlation functions that are, of course, contact term contributions. It is not obvious what role they play but it would be interesting to further explore their consequences in the future.
\subsection*{Correlators involving the scalar operator}
\subsection*{One scalar}
Let us first consider a correlator with a single scalar insertion and two currents. In the QF theory, such correlators are given by,
\begin{align}
    \langle J_{s_1} J_{s_2} O_2\rangle&=\frac{1}{\sqrt{\Tilde{N}}}\big(\langle J_{s_1} J_{s_2}  O_2\rangle_{FF}+\Tilde{\lambda}\langle J_{s_1} J_{s_2}  O_2\rangle_{CB}\big)\notag\\
    &=\frac{1}{\sqrt{\Tilde{N}}}\big(\langle J_{s_1} J_{s_2}  O_2\rangle_{FF}+\Tilde{\lambda}\epsilon\cdot \langle J_{s_1} J_{s_2}  O_2\rangle_{FF}\big)\,,
\end{align}
where in the second line, the epsilon transform is performed with respect to the lower amongst the two spins. This correlator in the four helicity configurations (after the rescaling \eqref{rescaling} and assuming $s_1\ge s_2$) is given by,
\begin{align}
    &\langle \Tilde{J}_{s_1}^{-}\Tilde{J}_{s_2}^{-}\Tilde{O}_2\rangle=\frac{e^{i\theta}}{\sqrt{\Tilde{N}}}\langle J_{s_1}^{-}J_{s_2}^{-}O_2\rangle_{FF},\notag\\&\langle \Tilde{J}_{s_1}^{+}\Tilde{J}_{s_2}^{+}\Tilde{O}_2\rangle=\frac{e^{-i\theta}}{\sqrt{\Tilde{N}}}\langle J_{s_1}^{+}J_{s_2}^{+}O_2\rangle_{FF},\notag\\
    & \langle \Tilde{J}_{s_1}^{-}\Tilde{J}_{s_2}^{+}\Tilde{O}_2\rangle=\frac{e^{i\theta}}{\sqrt{\Tilde{N}}}\langle J_{s_1}^{-}J_{s_2}^{+}O_2\rangle_{FF},\notag\\& \langle \Tilde{J}_{s_1}^{+}\Tilde{J}_{s_2}^{-}\Tilde{O}_2\rangle=\frac{e^{-i\theta}}{\sqrt{\Tilde{N}}}\langle J_{s_1}^{+}J_{s_2}^{-}O_2\rangle_{FF}\,.
\end{align}
Taking the anti-chiral limit \eqref{anti-chirallimit} we obtain,
\begin{align}
    &\langle \Tilde{J}_{s_1}^{-}\Tilde{J}_{s_2}^{-}\Tilde{O}_2\rangle=g_{ac}\langle J_{s_1}^{-}J_{s_2}^{-}O_2\rangle_{FF}\,,&&\langle \Tilde{J}_{s_1}^{+}\Tilde{J}_{s_2}^{+}\Tilde{O}_2\rangle=0\notag\\
    & \langle \Tilde{J}_{s_1}^{-}\Tilde{J}_{s_2}^{+}\Tilde{O}_2\rangle=g_{ac}\langle J_{s_1}^{-}J_{s_2}^{+}O_2\rangle_{FF}\,,&& \langle \Tilde{J}_{s_1}^{+}\Tilde{J}_{s_2}^{-}\Tilde{O}_2\rangle=0\,,
\end{align}
which is indeed the desired result.

\subsection*{Two scalars}
The QF theory correlator involving two scalars is given by
\begin{align}
    \langle J_{s}O_2O_2\rangle=\frac{(1+\Tilde{\lambda}^2)}{\sqrt{\Tilde{N}}}\langle J_{s}O_2O_2\rangle_{FF}\,.
\end{align}
In the two independent helicity configurations, it is given by,
\small
\begin{align}
    &\langle \Tilde{J}_{s}^{-}\Tilde{O}_2\Tilde{O}_2\rangle=\frac{e^{i\theta}}{\sqrt{\Tilde{N}}}\langle J_{s}^{-}\Tilde{O}_2\Tilde{O}_2\rangle_{FF},\langle \Tilde{J}_{s}^{+}\Tilde{O}_2\Tilde{O}_2\rangle=\frac{e^{-i\theta}}{\sqrt{\Tilde{N}}}\langle J_{s}^{+}O_2O_2\rangle_{FF}\,.
\end{align}
\normalsize
Taking the anti-chiral limit \eqref{anti-chirallimit} yet again results in the desired answer:
\begin{align}
    &\langle \Tilde{J}_{s}^{-}\Tilde{O}_2\Tilde{O}_2\rangle=g_{ac}\langle J_{s}^{-}O_2O_2\rangle_{FF}\,,&&\langle \Tilde{J}_{s}^{+}\Tilde{O}_2\Tilde{O}_2\rangle=0\,.
\end{align}
\subsection*{Three-point Summary}
Based on the examples considered above, one can easily show that the anti-chiral limit, \eqref{anti-chirallimit}, works for any three-point correlator that is inside or outside the triangle, also for the ones involving scalar operators, thus establishing that what we have uncovered is indeed the subsector that is dual to anti-Chiral higher-spin gravity. Indeed, at the three-point level, the main signature of (anti-)Chiral higher-spin gravity is that it features cubic interactions with the total helicity being (negative) positive.

Let us now explore the consequences of taking this limit at higher points, starting with the four-point case.

\section{Beyond three-point}
\label{sec:beyond}
A general comment is that beyond three-point and at the loop level chiral and anti-chiral interactions can mix in the bulk, see section \ref{sec:holo}, which is not in contradiction with having them as closed subsectors on their own. However, we should not expect the simple limiting procedure we proposed above to work as it is, but it is instructive to see what happens when we take the limit.

\subsection{Four-point functions}
In this subsection, we investigate what happens to four-point correlators in the QF theory when we take the anti-chiral limit \eqref{anti-chirallimit}. Let us begin with the scalar case.
\subsection*{\textbf{Example 1:~}$\langle O_2O_2O_2O_2\rangle$}
The scalar correlator in the CS+fermionic matter theory is given by \cite{Bedhotiya:2015uga,Turiaci:2018nua}
\small
\begin{align}
    \langle \Tilde{O}_2\Tilde{O}_2\Tilde{O}_2\Tilde{O}_2\rangle=\frac{\cos^4(\theta)(1+\Tilde{\lambda}^2)^2}{\Tilde{N}}\langle O_2O_2O_2O_2\rangle_{FF}=\frac{1}{\Tilde{N}}\langle O_2O_2O_2O_2\rangle_{FF},
\end{align}
\normalsize
where we used $\Tilde{\lambda}=\tan(\theta)$.
Therefore, in the chiral limit and the anti-chiral limits \eqref{chirallimit} and \eqref{anti-chirallimit}, this correlator goes to zero.\footnote{Given the fact that the fermionic scalar four-point function is non-local \cite{Sleight:2017pcz,Jain:2023juk}, it is a nice feature that it drops out in this limit. This is an indication that we are focusing on a local sub-sector.}

\subsection*{\textbf{Example 2:~}$\langle J_s O_2O_2O_2\rangle$}
This correlator in the two helicity configurations is given by \cite{Li:2019twz,Silva:2021ece}
\small
\begin{align}
    &\langle \Tilde{J}_s^{-}\Tilde{O}_2\Tilde{O}_2\Tilde{O}_2\rangle=\frac{e^{i\theta}}{\Tilde{N}}\big(\cos(\theta)\langle J_s^{-}O_2O_2O_2\rangle_{FF}+\sin(\theta)\langle J_s^{-}O_2O_2O_2\rangle_{CB}\big)\notag\,,\\
    &\langle \Tilde{J}_s^{+}\Tilde{O}_2\Tilde{O}_2\Tilde{O}_2\rangle=\frac{e^{-i\theta}}{\Tilde{N}}\big(\cos(\theta)\langle J_s^{+}O_2O_2O_2\rangle_{FF}+\sin(\theta)\langle J_s^{+}O_2O_2O_2\rangle_{CB}\big)\,.
\end{align}
\normalsize
If we now take the anti-chiral limit \eqref{anti-chirallimit} we see that
\begin{align}\label{JsO2O2O2aclimit}
    &\langle \Tilde{J}_s^{-}\Tilde{O}_2\Tilde{O}_2\Tilde{O}_2\rangle=\frac{g_{ac}^2}{2}\big(\langle J_s^{-}O_2O_2O_2\rangle_{FF}-i\langle J_s^{-}O_2O_2O_2\rangle_{CB}\big)\notag\\&\langle \Tilde{J}_s^{+}O_2O_2O_2\rangle=0\,,
\end{align}
thus showing that the anti-chiral limit \eqref{anti-chirallimit} is well defined and indeed picks out just the anti-chiral sector for these types of four-point functions.

In the above examples, we notice a very nice separation between the chiral (net negative helicity) and anti-chiral (net positive helicity) sectors. However, by investigating several more general four-point functions in appendix \ref{appendix:JJJJ}, we see that in addition to taking the (anti-)chiral limit, one needs to throw away some extra pieces to obtain finite results. However, we postpone further analysis at the four- and higher-point levels to future work.

\subsection{Anomalous dimensions in the (anti-)chiral limit}\label{sec:AnomalousDimensions}
The anomalous dimensions of the spin $s$ currents in the CS+fermionic matter theory at leading order in $\Tilde{N}$ are given by \cite{Giombi:2016zwa}
\small
\begin{align}
    \gamma_s=\frac{1}{\Tilde{N}}\bigg(\frac{a_s \Tilde{\lambda}^2}{(1+\Tilde{\lambda}^2)}+\frac{b_s \Tilde{\lambda}^2}{(1+\Tilde{\lambda}^2)^2}\bigg)=\frac{\sin^2(\theta)}{\Tilde{N}}\bigg(a_s+b_s \cos^2(\theta)\bigg)\,,
\end{align}
\normalsize
where we used $\Tilde{\lambda}=\tan(\theta)$. Let us now take the anti-chiral limit \eqref{anti-chirallimit}. In this limit we have
\begin{align}\label{anomalousdim}
\gamma_s=-\frac{a_s}{4}g_{ac}^2-\frac{b_s}{16}e^{2i\theta}g_{ac}^2+\cdots.
\end{align}
Note that $a_s$ is a rational function of $s$ that approaches constant for large spins, while $b_s\sim \log s $ for large spins. The latter signals that the CFT is a gauge theory. It seems that the chiral limit should freeze these gauge degrees of freedom and the $b_s$ term needs to be dropped.

As for the scalar operator, its anomalous dimension is given by \cite{Jain:2019fja}
\begin{align}
    \gamma_0=-\frac{32}{3\pi^2}\frac{\Tilde{\lambda}^2}{\Tilde{N}(1+\Tilde{\lambda}^2)}\,.
\end{align}
Using the definition $\Tilde{\lambda}=\tan\theta$ we see that this equals
\begin{align}
    \gamma_0=-\frac{32}{3\pi^2}\frac{(\sin\theta)^2}{\Tilde{N}}\,.
\end{align}
In the anti-chiral limit, this quantity is finite and equals,
\begin{align}\label{scalaranomalousdim}
    \gamma_0=\frac{8g_{ac}^2}{3\pi^2}\,.
\end{align}
An important point to note here is that \eqref{anomalousdim} and \eqref{scalaranomalousdim} could receive corrections from non-planar diagrams. In the usual setting, we take the 't Hooft limit $\lambda=\frac{N}{k}$ to be finite while taking $N$ and $k$ to infinity. However, the anti-chiral limit \eqref{anti-chirallimit} that we take corresponds to $\lambda\to -i\infty$ which means we take $N\to \infty$ faster than we take $k\to\infty$. This can alternatively be interpreted as $N\to \infty$ while taking $k\to i$ thus indicating that the contributions of the non-planar diagrams to the anomalous dimensions need to be kept while taking this limit. Thus, it may be appropriate or even necessary to perform the re-summation of these contributions. For instance, the divergent part of the anomalous dimension of the spin $s$ currents \eqref{anomalousdim} could be rendered finite by these contributions. We leave a detailed investigation of this to the future.

\section{Holographic predictions for correlators}
\label{sec:holo}
In this section we change the gears and move into the bulk. It is quite easy to get a number of predictions by analysing the structure of interactions in Chiral theory without having to actually compute any holographic correlators. Some of the statements below go a bit beyond what is known at present on the CFT side. Several points should be stressed before we proceed. 

Firstly, as we show below, certain corrections vanish identically in the bulk because it is impossible to draw a tree/loop diagram that would contribute. These are robust predictions. However, we expect that the higher-spin symmetry implies many cancellations. For example, there have to be no loop corrections to holographic correlators for the free CFT duals. Therefore, certain combinations of loop diagrams must vanish, even though it is possible to draw them. To take this possibility into account to refer to a situation when a contribution exists, but only an explicit computation can tell us if it vanishes or not, we will use the sign $\bulkneq0$. 

Secondly, we will consider the most general case of the bosonic ($U(M)$ or $O(M)$) gauged Chiral higher-spin gravity. On the CFT side, this corresponds to vector models with some leftover global symmetries, e.g. $U(M)$ or $O(M)$.\footnote{The $O(N)$-gauging relevant for Chiral higher-spin gravity was first discussed in \cite{Metsaev:1991nb} and its generalization in \cite{Skvortsov:2020wtf}, which is very similar to \cite{Konstein:1989ij}. } However, there are two standard cases in the literature:\footnote{The two cases correspond to Chern-Simons matter theories with $U(N)$ and $O(N)$ gauged symmetry and without the leftover $U(M)$ or $O(M)$ symmetry.} (a) the spectrum contains higher-spin currents of all integer spins; (b) the spectrum has higher-spin currents with even spins only. These are obtained by simple reductions of the general statements that are made for the nonabelian Chiral theory: for (a) the sum of spins meeting at a vertex has to be even; for (b) the spins themselves have to be even. For example, for case (a) $\langle TTJ\rangle=0$ and there is no bulk cubic vertex with spins $2-2-1$. Supersymmetric extensions are also possible, but we do not discuss them in detail.

Thirdly, one should be careful in comparing the bulk predictions to CFT while the chiral limit on the CFT side is not yet well-understood. Indeed, the hypothetical bulk dual of the full Chern-Simons matter theories (see also the picture in the introduction) has the same spectrum as Chiral theory. Chiral and anti-chiral interactions cover all possible cubic interactions. Starting from the quartic order this is not the case and there has to be some structures (let us call them nonchiral) not belonging to (anti-)Chiral theories. Starting from four-point functions (and loop corrections) holographic correlators receive contributions from (anti-)Chiral theories separately and from various mixtures of (anti-)chiral interactions together with the nonchiral ones.    

Let us begin by giving a schematic form of the Lagrangian of Chiral theory. The Lagrangian is known in the light-cone gauge in flat space \cite{Metsaev:1991nb,Metsaev:1991mt,Ponomarev:2016lrm}, up to the cubic order in $AdS_4$ \cite{Metsaev:2018xip,Skvortsov:2018uru}. With the equations of motion \cite{Skvortsov:2018uru,Sharapov:2022wpz,Sharapov:2022awp} it is easy to see what kind of contact interaction vertices are present at higher orders in $AdS_4$, which can be summarized as 
\begin{align}
    \mathcal{L}&= \sum_{s\geq0} \Phi_{-s} \square \Phi_{s}+\sum_{N\geq3} g^{N-2} \sum_{\lambdatot\geq N-2} V^N_{\lambda_1,...,\lambda_N}\,.
\end{align}
Here, the first term is the sum of kinetic terms for all spins. Note that a massless spin-$s$, $s>0$ field has two degrees of freedom that can be associated with helicity $+s$ and $-s$ states, which we denote $\Phi_{\pm s}$.\footnote{The Lagrangian is written as if we use the light-cone gauge \cite{Metsaev:2018xip} where the helicity structure is manifest. Note that the kinetic terms do reduce to simple $\square=\partial_m \partial^m$ in the light-cone gauge after an appropriate rescaling despite $AdS_4$ is curved, which is due to the fact that massless higher-spin fields are also conformally invariant in certain field descriptions.} In particular, the bulk-to-bulk propagator connects helicity $\lambda$ to helicity $-\lambda$. The most important information about the interactions is that at order $N$ the total helicity $\lambdatot=\sum_i\lambda_i$ entering the vertex $V^N_{\lambda_1,...,\lambda_N}$ must be greater or equal to $N-2$ for chiral theory or $-\lambdatot>N-2$ for the anti-Chiral one. For example, all cubic interactions with $\lambda_1+\lambda_2+
\lambda_3>0$ are present in Chiral theory and those with $\lambda_1+\lambda_2+
\lambda_3<0$ are present in anti-Chiral one. In reality, the helicity is integer in the bosonic theory and we have $\lambda_1+\lambda_2+
\lambda_3\geq1$ (or $\leq-1$). For simplicity, we confine ourselves to anti-Chiral theory as in the main text above.\footnote{Sketchy arguments to explain the possibility of the chiral truncation from the bulk vantage point can be found in appendix \ref{app:bulklimit}. }

\paragraph{Three-point functions.} The main constraint for the three-point functions is the aforementioned $\lambda_1+\lambda_2+
\lambda_3<0$. In particular, $0-0-0$ vanishes. Therefore, $\langle OOO\rangle=0$ and all three-point correlators where the sum of the helicities greater or equal to zero vanish. This is exactly what we observed above. It is important that Chiral theory can be shown to be a unique theory under certain assumptions.\footnote{Assumptions can vary, for example, one can assume that there is at least one higher-spin field that has nontrivial self-interactions. The latter forces one to introduce all other spins and all other (anti-)chiral interactions, see \cite{Metsaev:1991nb,Metsaev:1991mt,Ponomarev:2016lrm}. It is important to note that there are two contractions of Chiral theory \cite{Ponomarev:2017nrr} that can be thought of as higher-spin extensions of selfdual Yang-Mills and of self-dual gravity (separately). They contain higher-spin fields but exhibit only one-derivative Yang-Mills-type interactions or two-derivative gravitational-type interactions, but no genuine higher-spin interactions. These two theories have simple actions \cite{Krasnov:2021nsq}. }

\paragraph{Four-point functions.} There are two types of contributions to the four-point functions at tree level: exchanges and contact, but they satisfy the same constraint $-\lambdatot\geq N-2$ as one can easily see. Therefore, for a four-point correlator to be nonzero we should have $\lambda_1+\lambda_2+\lambda_3+\lambda_4\leq-2$. 

For example, for $0-0-0-0$ the only vertex that can contribute to the exchange is $0-0-(-s)$ and the bulk-to-bulk propagator connects $+s$ to $-s$ and $0-0-(\pm s)$ cannot be both in the same Chiral theory. Therefore, $\langle OOOO\rangle=0$ and, in general, we expect
\begin{align}
    \langle J_{-s} OOO\rangle & \bulkneq0\,, \quad s\geq2\,, & \begin{aligned}
        \langle J_{-s} OOO\rangle & =0\,, \quad s=0,1\,,\\
        \langle J_{+s} OOO\rangle & =0\,, \quad s\geq0\,,
    \end{aligned} 
\end{align}
where the correlator on the left can still be zero as a result of some cancellation, but in general we expect that it does not vanish.

\paragraph{Higher-point functions.} For higher point functions we have $-\lambdatot\geq N-2$. In particular, we expect that 
\small
\begin{align}
        \langle J_{-s} OOOO\rangle & \bulkneq0\,, \quad s\geq3\,, &
    \begin{aligned}
        \langle J_{-s} OOOO\rangle & =0\,, \quad s=0,1,2\,,\\
        \langle J_{+s} OOOO\rangle & =0\,, \quad s\geq0
    \end{aligned} 
\end{align}
\normalsize
For example, all $\langle OO...O\rangle=0$.

\paragraph{Loop corrections.} Flat space computations \cite{Skvortsov:2018jea,Skvortsov:2020gpn} show that there are no UV divergences in the theory at least at one-loop. Any $n$-point one-loop diagram can be obtained by connecting two lines in some $(n+2)$-point tree-level diagram. Therefore, for a one-loop correction to an $n$-point correlator we get $-\lambdatot\geq n$. 

For example, for two-point functions, we have $\lambda_1+\lambda_2\leq-2$. Therefore, we do not expect any one-loop correction to $\langle OO \rangle$, but there can be one to $\langle J_{-s} J_{-s}\rangle$, $s\geq1$. For three-point functions at one-loop, we have $\lambda_1+\lambda_2+
\lambda_3\leq -3$ and so on. In particular, there should not be any loop corrections to $\langle OO...O\rangle$. 

Assuming Chiral theory does not have any UV-divergences to all orders and, hence, no new counterterms need to be introduced that could change the simple conclusions here, we can estimate that an $l$-loop correction to an $n$-point function comes by gluing $2l$ lines in a $(2l+n)$-point tree level diagram. Therefore, we have to have $-\lambdatot\geq 2l+n-2$. As a consequence, we see that for any given correlation function $\langle J_{\lambda_1} ...J_{\lambda_n}\rangle$ loop corrections stop at a certain finite loop order, $2l\leq -\lambdatot-n+2$. The latter means that the $1/N$ expansion is convergent (we can say nonuniformly because the order depends on the observable being computed).

\section{Discussion and Conclusions}
\label{sec:conc}
In the paper, we made the first step towards identifying the hidden closed subsector of Chern-Simons matter theories that is dual to Chiral higher-spin gravity. It is quite easy to do so at the level of three-point functions by taking a certain limit of the coupling constant. The main selection rule here is to keep the correlators with total helicity positive/negative. At higher orders the picture is less clear since chiral and anti-chiral sectors can mix. Indeed, this is what we observed in appendix \ref{appendix:ExtraFourPoint}. However, if we naively attempt to bootstrap higher-point correlators of the (anti-)chiral theory using the three-point functions as an input, we obtain simple but different results from directly taking the (anti-)chiral limit of the correlators as we showed in appendix \ref{appendix:CPW}. The latter procedure should be consistent with the bulk picture.

One should not rush to compare some of our bulk conclusions to the CFT results above because the existence of a closed subsector is not the same as existence of a limiting subsector. We see that at higher orders chiral and anti-chiral theories/subsectors mix with each other and with the nonchiral ones and a further refinement of our limiting procedure may be needed. For example, one may need to drop certain parts of correlation functions to land on the chiral subsector. 

It is worth stressing that the perturbation series in the bulk seems to converge (nonuniformly in spins), which is dual to the convergence of the $1/N$ expansion in the chiral subsector. The chiral subsector can be bootstrapped with the help of the slightly-broken higher-spin symmetry starting from the three-point functions we have. Therefore, there is a good chance of establishing a pair of AdS/CFT dual theories where both sides are nontrivial, computable and are defined independently of each other. 

In general, one might expect the slightly-broken higher-spin symmetry to be useful even at small $N$, given that the anomalous dimensions of higher-spin currents are small and get even smaller with spin. It would be interesting to find out if the existence of chiral subsectors can also be established directly at small $N$ via the numerical bootstrap techniques, where the main challenge seems to be to implement the notion of helicity into the structure of correlators and OPE coefficients and to be able to target nonunitary theories.

On the way towards exact models of AdS/CFT correspondence, it is interesting to identify the AdS/CFT duals of self-dual Yang-Mills and of self-dual gravity. In flat space all tree-level amplitudes in these theories vanish and the only nontrivial ones are all-helicity-plus one-loop amplitudes. However, it is not so in (anti-)de Sitter and it would be interesting to compute the corresponding AdS/CFT correlators. Already tree-level amplitudes of type $(++...+-)$ do not vanish. The existence of these theories implies that there are closed subsectors of the current $\langle J...J\rangle$ and stress tensor $\langle T...T\rangle$ correlators. However, these two subsectors are quite small. Nevertheless, they are universal since every (local) CFT has a stress tensor and many have global symmetry currents, while the existence of the slightly-broken higher-spin currents is special to large-$N$ vector models. It also should be taken into account that both SDYM and SDGR are UV finite theories \cite{Krasnov:2016emc}. 

SDYM and SDGR are contained inside (anti-)Chiral theory. SDYM has only the $(--+)$ cubic vertex and SDGR also has the $(--+)$ vertex and stops there at least in the light-cone gauge \cite{Lipstein:2023pih,Neiman:2023bkq}. Therefore, our expressions for $\langle T^-T^-T^+\rangle_{nh}$ and $\langle J^-J^-J^+\rangle_{nh}$ immediately give the three-point holographic functions in SDGR and SDYM, respectively. As for the four-point ones, it easy to see that $\langle J^-J^-J^-J^+\rangle $ in SDYM coincides with the one in Chiral theory with all integer spins. Similarly, $\langle T^-T^-T^-T^+\rangle $ in SDGR coincides with the one on Chiral theory restricted to even spins. In fact the same should be true for all $\langle J^-J^-...J^-J^+\rangle $ and $\langle T^-T^-...T^-T^+\rangle $ since exchanges via higher-spin fields are suppressed in these correlators.

In the same vein, it would be interesting to identify the chiral subsectors in ABJ theories since via triality \cite{Chang:2012kt} one can argue that $\mathcal{N}=6$ $U(M)$-gauged Chiral theory should be dual to the subsector of the vector-like limit of ABJ theories. The latter indicates the existence of a closed subsector in tensionless strings on $AdS_4\times \mathbb{CP}^3$.

Our (anti-)chiral limit can be understood as taking $N\to\infty$ while keeping the Chern-Simons level fixed at $k=\pm i$. The analysis of Chern-Simons theory with a complex level has been performed by Witten \cite{Witten:1989ip,Witten:2010cx}. It would be interesting to see if one can perform an analogous analysis when we also add matter into the mix and see for instance, if the bosonization dualities persist.
In the literature, supersymmetric Chern-Simons matter theories have been analyzed in the so-called M-theory limit which corresponds to $N\to\infty$ with $k$ finite \cite{Kapustin:2009kz,Marino:2011eh}, which is quite similar in spirit to our (anti-)chiral limit. By localization, the partition function on $S^3$ reduces to a matrix model which is then interpreted as the partition function of an ideal Fermi gas of non-interacting particles. The level $k$ is mapped to Planck's constant and, hence, one can study the theory at finite $k$ by employing the WKB approximation. This formalism also enables one to compute non-perturbative (instanton) effects. It would thus be interesting to understand the implications of the (anti-)chiral limits in this vein.

The works mentioned above were in the context of ABJM or ABJ theory with matter in the adjoint or bi-fundamental representations. There has also been some work on the supersymmetric versions of Chern-Simons theory with fundamental matter \cite{Inbasekar:2015tsa,Gur-Ari:2015pca,Aharony:2019mbc}. It would be interesting to see if we can define (anti-)chiral limits analogous to \eqref{chirallimit} and \eqref{anti-chirallimit} to obtain a sub-sector that is dual to a supersymmetric (anti-)chiral higher-spin gravity. The formalism of super spinor-helicity variables developed in \cite{Jain:2023idr} for $\mathcal{N}=1$ and $\mathcal{N}=2$ theories would provide a perfect stage for the start of such exploits.

Another avenue of interest is the connection between $\text{AdS}_4/\text{CFT}_3$ and flat space holography. $\text{CFT}_3$ correlators/ $\text{AdS}_4$ amplitudes are related to S-matrix in four dimensions. The framework of Celestial holography relates four-dimensional S-matrices to $\text{CFT}_2$ correlators. It would thus be interesting to see the implications of taking the (anti-)chiral limit on the celestial correlators.

\section*{Acknowledgments}
\label{sec:Aknowledgements}
We would like to thank Ofer Aharony, Trivko Kukolj and Rohit R. Kalloor for useful discussions. We also acknowledge that while working on this paper, we became aware of \cite{Aharony:2024nqs} that has an overlap with the main points of this work. We are also grateful to the authors of \cite{Aharony:2024nqs} for coordinating the submissions of the pre-print versions of our papers to arxiv. We are grateful to Simone Giombi, Saurabh Pant and Sasha Zhiboedov for very useful discussions. SJ would also like to thank the organizers and participants of conferences organized in BITS Goa and IIT Kanpur where part of this work was presented. This project has received funding from the European Research Council (ERC) under the European Union’s Horizon 2020 research and innovation programme (grant agreement No 101002551). E.S. is a Research Associate of the Fund for Scientific Research -- FNRS, Belgium. S.J and D.K.S would like to acknowledge their debt to the people of India for their steady support of research in basic sciences.

\appendix

\clearpage
\onecolumngrid
\section*{APPENDIX}
\section{Holographic Chiral limit}
\label{app:bulklimit}
Let us discuss the chiral limit in the bulk, in the holographic dual of Chern-Simons vector models. This theory is understood at present only at the level of the underlying $L_\infty$-algebra. This algebra determines, in principle, formally consistent classical equations of motion in the form of a free differential algebra \cite{Vasiliev:1990en,Vasiliev:1992av}, see also \cite{Sharapov:2019vyd} for the general solution to the problem of constructing $L_\infty$-extension starting from free higher spin fields. We will not need any concrete technical details of the construction to identify the chiral limit. There are two fields: a one-form $\omega$ and a zero-form $C$, which are generating functions containing fields of all spins together with certain auxiliary fields. Based on the form degree counting, the general structure has to be
\begin{align}
    d\omega&= \mathcal{V}(\omega,\omega)+\mathcal{V}(\omega,\omega,C)+\mathcal{V}(\omega,\omega,C,C) +\mathcal{O}(C^3)\,,\\
    dC&= \mathcal{U}(\omega,C)+\mathcal{U}(\omega,C,C)+\mathcal{O}(C^3)\,,
\end{align}
where $\mathcal{V}$ and $\mathcal{U}$ are certain multilinear maps (the structure maps of some $L_\infty$-algebra). These multilinear maps have the following general structure
\begin{align}
    \mathcal{V}(\omega,\omega,C)&= g\,\mathcal{V}_{10}(\omega,\omega,C)+\bar{g}\,\mathcal{V}_{01}(\omega,\omega,C)\,,\\
    \mathcal{V}(\omega,\omega,C,C)&=g^2\,\mathcal{V}(\omega,\omega,C,C)_{20}+\bar{g}g\,\mathcal{V}_{11}(\omega,\omega,C,C)+\bar{g}^2\,\mathcal{V}_{02}(\omega,\omega,C,C)\,,\\
    \mathcal{U}(\omega,C,C)&=g\,\mathcal{U}_{10}(\omega,C,C)+\bar{g}\,\mathcal{U}_{01}(\omega,C,C)\,,
\end{align}
where $g$ and $\bar{g}$ can be taken to be independent complex coupling constants. More generally, we have
\begin{align}
    \mathcal{V}(\omega,\omega,\overbrace{C,...,C}^n)&= \sum_{i+j=n}g^i \bar{g}^j \,\mathcal{V}_{i,j}(\omega,\omega,\overbrace{C,...,C}^n)\\
    \mathcal{U}(\omega,\overbrace{C,...,C}^{n+1})&= \sum_{i+j=n}g^i \bar{g}^j \,\mathcal{U}_{i,j}(\omega,\overbrace{C,...,C}^{n+1})
\end{align}
The general structures above can be either extracted from Vasiliev's equations \cite{Vasiliev:1990en,Vasiliev:1992av}, or defined directly \cite{Sharapov:2019vyd}. The $L_\infty$-algebra itself is defined only modulo canonical transformations, which correspond to field redefinitions 
\begin{align}
    \omega&\rightarrow \omega+f(\omega,C) & C&\rightarrow C+ g(C)
\end{align}
from the field theory point of view. Such redefinitions are in general very nonlocal from the field theory point of view and it is still an open problem to define the vertices $\mathcal{V}$ and $\mathcal{U}$ in such a way that systematic holographic calculations are possible. Assuming that the resulting theory can be defined and the locality considerations do not modify the general structure of the equations,\footnote{Note that $g$ and $\bar{g}$ correspond to two nontrivial deformations at the $L_\infty$-level \cite{Vasiliev:1990en,Vasiliev:1992av}, see also \cite{Sharapov:2020quq}. In fact, provided locality is not taken into account, there are infinitely many other deformations that enter at each odd order in $C$, i.e. in $\mathcal{V}_{i,j}$ with $i+j=\text{odd}$. They correspond to the phase function $\theta_\star(X)=-\theta_\star(-X)$ in Vasiliev's equations, \cite{Vasiliev:1992av}. These deformations do not seem to find any natural explanation on the CFT side, see e.g. \cite{Maldacena:2012sf,Boulanger:2015ova}. It may well be that the holographic dual of Chern-Simons vector models may need to activate (some of) these parameters, see e.g. \cite{Giombi:2011kc}. This will complicate the matters discussed here. Another possibility is that the careful treatment of locality can change the cohomological analysis, i.e. the structure of the deformation, by introducing some new/eliminating some old deformation parameters. We assume that none of this will happen. } it is easy to see that the (anti-)chiral limits should correspond simply to $g=0$ or $\bar{g}=0$. Imposing unitarity one has to have $g=|g|e^{i\theta}$ and $\bar{g}=|g|e^{-i\theta}$, where $|g|\sim 1/\sqrt{N}$ is the bulk coupling. Therefore, the chiral limit is nonunitary. Historically, the possibility of existence of self-dual higher-spin gravities was entertained already in \cite{Vasiliev:1992av} and the present discussion at the level of formally consistent equations should be equivalent to \cite{Vasiliev:1992av}.

From the bulk perspective, the chiral limit has improved locality properties, and the perturbatively local equations of motion are known explicitly \cite{Sharapov:2022wpz,Sharapov:2022awp} and are a smooth uplift of the Chiral theory from flat space \cite{Skvortsov:2018uru} to anti-de Sitter.

\section{Notation and three dimensional spinor-helicity variables}\label{appendix:Notation}
In this appendix, we set straight the notation used in the paper. The abbreviations we use are provided in the table below.
\begin{table}[h!]\label{tab:table1}
  \begin{center}
    \begin{tabular}{|c|l|}
    \hline    
      \textbf{Abbreviation} & \textbf{Full Form} \\
      \hline
      $\langle ...\rangle_\text{QF}$ & In quasi-fermionic theory\\
      \hline
      $\langle ...\rangle_\text{FF/FB}$ & In free fermionic/bosonic theory\\
      \hline
      $\langle ...\rangle_\text{CB}$ & In critical bosonic theory\\
      \hline
      $\langle ...\rangle_{\text{odd}}$ & Parity odd correlator\\
      \hline
      $\langle ...\rangle_\text{FF+FB}$ & $\langle ...\rangle_\text{FF}+\langle ...\rangle_\text{FB}$\\
      \hline
      $\langle ...\rangle_\text{FF-FB}$ & $\langle ...\rangle_\text{FF}-\langle ...\rangle_\text{FB}$\\
      \hline
      $\langle ...\rangle_\text{nh}$ & nonhomogeneous\\
      \hline
      $\langle ...\rangle_\text{h}$ & homogeneous\\
      \hline
    \end{tabular}
    \caption{Some notation}
  \end{center}
\end{table}
While writing correlators in momentum space/spinor-helicity variables we suppress the momentum-conserving delta functions.
In spinor-helicity variables, we denote the negative (positive) helicity component of a spin $s$ current $J_s^{\mu_1\cdots \mu_s}$ that one obtains after contraction with the transverse null polarizations $z_{\mu_1}^{\pm}\cdots z_{\mu_s}^{\pm}$ as $J_s^{-}$($J_s^{+}$). This is shorthand for $J_s^{-\cdots -}$ ($J_s^{+\cdots+}$) where there are $s$ minuses (pluses). We now briefly review the formalism of three-dimensional spinor-helicity variables.

We work with the usual flat Euclidean metric,
\begin{align}
    \delta_{\mu\nu}=\textbf{diag}(1,1,1),
\end{align}
with which vector indices are raised and lowered. Since upper and lower indices are identical in this case, we do not need to distinguish between them. Our convention for the three-dimensional Levi-Civita symbol is,
\begin{align}
\epsilon_{123}&=\epsilon^{123}=1,\notag\\
    \epsilon_{\mu\nu\rho}\epsilon_{\alpha\beta\rho}&=\delta_{\mu\alpha}\delta_{\nu\beta}-\delta_{\mu\beta}\delta_{\nu\alpha}.
\end{align}
Spinor indices on the other hand are raised and lowered using the two-dimensional Levi-Civita symbol $\epsilon_{ab}$ which is given by,
\begin{align}
\epsilon_{12}&=\epsilon^{12}=1,\notag\\
    \epsilon_{ab}\epsilon^{ac}&=\delta_b^c.
\end{align}
Our convention for raising and lowering indices is as follows: given a spinor $A_a$ we have,
\begin{align}\label{IndexRaisingSpinor}
    A^a=\epsilon^{ab}A_b\iff A_a=\epsilon_{ba}A^b.
\end{align}
We choose the following representations of the Pauli matrices 
\begin{align}
    &\sigma^1=\begin{pmatrix}0&1\\
    1&0\end{pmatrix}\,,\qquad\sigma^2=\begin{pmatrix}0&-i\\
    i&0\end{pmatrix}\,,\qquad\sigma^3=\begin{pmatrix}1&0\\
    0&-1\end{pmatrix}\,,
\end{align}
which satisfy,
\begin{align}
    (\sigma^\mu)^a_b(\sigma^\nu)^c_a=\delta^{\mu\nu}\delta^c_b+i\epsilon^{\mu\nu\rho}(\sigma^\rho)^c_b.
\end{align}
Given a three-vector $p_\mu$, we can trade it for a matrix $(\slashed{p})^{a}_{b}$ in the following way:
\begin{align}
    (\slashed{p})^{a}_{b}=p_\mu(\sigma^\mu)^a_b=\lambda_b\Bar{\lambda}^a+p\delta^a_b.
\end{align}
We can extract the ``energy" through the following bracket:
\begin{equation}
    p=-\frac{1}{2}\langle \lambda \Bar{\lambda}\rangle.
\end{equation}
By contracting spinors associated to different momenta, we can form the angle bracket:
\begin{align}
    &\langle i j\rangle=\lambda_{ia}\lambda_j^a.
\end{align}
 The polarization vectors associated to the currents are given by,
\begin{align}
    &z_\mu^{-}(\sigma^\mu)^a_b=(\slashed{z}^{-})^a_b=\frac{\lambda_b\lambda^a}{p},\\
    &z_\mu^{+}(\sigma^\mu)^a_b=(\slashed{z}^{+})^a_b=\frac{\Bar{\lambda}_b\Bar{\lambda}^a}{p}.
\end{align}
Polarization vectors of higher-spin operators can be formed by taking products of these spin-one polarization vectors. Momentum conservation in these variables is given by,
\begin{align}\label{momconsSH}
    \sum_{i=1}^{n}\lambda_{ib}\Bar{\lambda}_i^a=-E\delta^a_b.
\end{align}
$E=\sum_{i=1}^{n}p_i$ is the total energy. The three-dimensional dot product of two three-vectors $x$ and $y$ can be written in spinor notation using,
\begin{align}
    &x\cdot y=\frac{1}{2}(\slashed{x})^a_b(\slashed{y})^b_a.
\end{align}
Since we work with parity odd correlation functions as well, we will require the following formula:
\begin{align}
&\epsilon^{\mu\nu\rho}=\frac{1}{2i}(\sigma^\mu)^a_b(\sigma^\nu)^c_a(\sigma^\rho)^b_c.
\end{align}
Finally, we have used the following notation:
\begin{equation}
    \epsilon^{v_1 v_2 v_3}=v_{1\mu}v_{2\nu}v_{3\rho}\epsilon^{\mu\nu\rho},
\end{equation}
for any three vectors $v_1,v_2$ and $v_3$. More details as well as some useful formulae can be found in appendix A of \cite{Jain:2023idr}.

\section{A Chiral limit for the Quasi Bosonic theory}\label{appendix:QBchirallimit}
In this appendix, we shall define a chiral limit for correlation functions in the Chern-Simons$+$fundamental bosons theory (henceforth called the quasi-bosonic (QB) theory). Let us first present the Lagrangian description of the theory. 

The $U(N)$ Regular bosonic theory is described by the following action:
\begin{align}
    S=\int d^3x~\bigg(D_\mu\Bar{\phi}D^\mu\phi+i\epsilon^{\mu\nu\rho}\frac{\kappa_b}{4\pi}\Tr{A_\mu\partial_\nu A_\rho-\frac{2i}{3}A_\mu A_\nu A_\rho}+(\frac{2\pi}{k_b})^2(\lambda_6+1)(\Bar{\phi}\phi)^3\bigg).
\end{align}
We take the limits $\kappa_b\to\infty,N\to\infty$ while keeping fixed $\lambda=\frac{N}{k_b}$. 

This theory enjoys a strong-weak duality with the critical fermionic theory, the action of which is provided in equation $(1.5)$ in \cite{Aharony:2018pjn}.

In terms of correlators, the main difference between the QF and QB theories is the fact that the three-point function of scalars is nonzero in the latter case. It is given by \cite{Aharony:2012nh}\footnote{Our normalization for the operator in contrast to section 5.1 in \cite{Aharony:2012nh} is as follows: $O_{1,ours}=\frac{O_{1,theirs}}{\sqrt{N(1+\Tilde{\lambda}^2)}}$. Also, we defined the quantity $\Tilde{\lambda}=\tan(\frac{\pi\lambda}{2})$.}
\begin{align}
    \langle O_1 O_1 O_1\rangle_{QB}=\frac{1}{\sqrt{N(1+\Tilde{\lambda}^2)}}\bigg(1+\frac{\Tilde{\lambda}^2}{4}\big(3-\frac{\lambda_6}{8\pi^2\lambda^2}\big)\bigg)\langle O_1O_1O_1\rangle_{FB}.
\end{align}
We see that we can make this three-point function identically zero by choosing the coefficient of the $\phi^6$ term, $\lambda_6$ as follows:
\begin{align}
    \lambda_6=32(3+\frac{4}{\Tilde{\lambda}^2})\tan^{-1}(\Tilde{\lambda})^2.
\end{align}
We then define chiral and anti-chiral limits much in the same way as \eqref{chirallimit} and \eqref{anti-chirallimit}. The analysis of the spinning correlators is then identical to that of the quasi-fermionic theory presented in section \ref{sec:ChiralHSfromCS} since the $\lambda_6$ coefficient only contributes to the scalar three-point function.

\section{Generalization of statements in section \ref{sec:CFT3corrfacts} to correlators involving fermionic operators}
\label{appendix:halfintcase}
In this appendix, we shall generalize our statements about correlation functions inside the triangle in section \ref{sec:CFT3corrfacts} (\eqref{statement1},\eqref{statement2} and \eqref{statement3}, to be precise) to those that include half-integer spin operators. The reason for the same is to show that our statements are general enough. For brevity, we shall only focus on the nonhomogeneous parts of the correlators. This is because the homogeneous contributions all obey statement $1$ \eqref{statement1}. Let us begin with a simple example.\footnote{The correlators that appear in this section can be computed from the free bosonic and free fermionic theories and using the definition of the nonhomogeneous piece \eqref{3ptCFThandnh}. Schematically, for the Abelian examples, $O_{1/2}\sim \psi \Bar{\phi}+cc$, $J_{3/2}\sim \psi\partial\Bar{\phi}+cc$ and $J$ and $T$ are the $U(1)$ current and the stress tensor respectively.}

\subsection*{\textbf{Example 1:}~$\langle O_{1/2}O_{1/2}J\rangle$}
The nonhomogeneous contribution to this correlator is nonzero in the $(- - -)$ and the zero-helicity sector $(- - +)$ and their complex conjugate ones.\footnote{These expressions can be obtained by solving the conformal Ward identities in spinor-helicity variables like in \cite{Jain:2021vrv}.} The expressions are respectively given by,
\begin{align}
    &\langle O_{1/2}^{-A}O_{1/2}^{-B}J^{-B}\rangle_{nh}=f^{ABC}\frac{\langle 2 3\rangle\langle 3 1\rangle}{\sqrt{p_1 p_2}p_3^2},\notag\\
    &\langle O_{1/2}^{-A}O_{1/2}^{-B}J^{+C}\rangle_{nh}=f^{ABC}\frac{\langle 1 2\rangle^2((p_1-p_2)^2-p_3^2)}{\langle 2 3\rangle\langle 3 1\rangle \sqrt{p_1 p_2}p_3^2}.
\end{align}
Consider now, the following contact term:
\begin{align}
    \langle O_{1/2}^{A}O_{1/2}^{B}J^{C}\rangle_{contact}=-f^{ABC}\frac{\zeta_{1a}\zeta_{2b}(\slashed{z}_3)^{ab}}{p_3},
\end{align}
and redefine,
\begin{align}
    \langle O_{1/2}^{A}O_{1/2}^{B}J^{C}\rangle_{new,nh}=\langle O_{1/2}^{A}O_{1/2}^{B}J^{C}\rangle_{nh}-\langle O_{1/2}^{A}O_{1/2}^{B}J^{C}\rangle_{contact}.
\end{align}
The redefined nonhomogeneous correlator satisfies,
\begin{align}
    \langle O_{1/2}^{-A}O_{1/2}^{-B}J^{-B}\rangle_{new,nh}=\langle O_{1/2}^{-A}O_{1/2}^{-B}J^{+B}\rangle_{new,nh}=0,
\end{align}
that is, it vanishes in the $(- - -)$ as well as zero helicity sector (and of course, their complex conjugate configurations).
\subsection*{\textbf{Example 2:}$~\langle J_{3/2}J_{3/2}J\rangle$}
Let us now focus on a correlator that involves two spin three-half currents. The nonhomogeneous part of this correlator in the $(- - -)$ helicity is given by,
\begin{align}
    \langle J_{3/2}^{-}J_{3/2}^{-}J^{-}\rangle_{nh}=\frac{\langle 1 2\rangle^2\langle 2 3\rangle\langle 3 1\rangle (2E+p_3)}{(p_1p_2)^{3/2}p_3}.
\end{align}
We can, however, cancel this contribution by adding the following simple contact term:
\begin{align}
     \zeta_{1a}\zeta_{1b}\zeta_{1c}\zeta_{2e}\zeta_{2f}\zeta_{2g}z_{3\mu}\langle J_{3/2}^{abc}J_{3/2}^{efg}J^{\mu}\rangle_{contact}=-\zeta_{1a}\zeta_{1b}\zeta_{1c}\zeta_{2e}\zeta_{2f}\zeta_{2g}z_{3}^{gc}\epsilon^{af}\epsilon^{be}(E+p_3).
\end{align}
Redefining
\begin{align}
     \langle J_{3/2}J_{3/2}J\rangle_{new,nh}=\langle J_{3/2}J_{3/2}J\rangle_{nh}-\langle J_{3/2}J_{3/2}J\rangle_{contact},
\end{align}
we find
\begin{align}
     \langle J_{3/2}^{-}J_{3/2}^{-}J^{-}\rangle_{new,nh}=0,
\end{align}
showing that the nonhomogeneous part of this correlator is zero when the helicities of all the insertions coincide.
\subsection*{\textbf{Example 3:}~$\langle J_{3/2}O_{1/2}J\rangle$}
The nonhomogeneous part of this correlator is nonzero in both the $(- - -)$ and $(+ - -)$ helicities which should not be the case according to statements $2$ and $3$ (\eqref{statement2} and \eqref{statement3}):
\begin{align}
    &\langle J_{3/2}^{-}O_{1/2}^{-}J^{-}\rangle=-\frac{\langle 1 2\rangle\langle 3 1\rangle^2}{p_1^{3/2}\sqrt{p_2}p_3}\notag\\
    &\langle J_{3/2}^{+}O_{1/2}^{-}J^{-}\rangle=\frac{\langle 2 3\rangle^3(E-2p_3)^2(E-2p_2)}{\langle 1 2\rangle^2\langle 3 1\rangle p_1^{3/2}\sqrt{p_2}p_3}.
\end{align}
However, these contributions can be removed via the addition of the following contact term:
\begin{align}
    z_{1abc}\zeta_{2b}z_{3\mu}\langle J^{abc}_{3/2}O_{1/2}^{b}J^{\mu}\rangle_{contact}=\frac{2\zeta_{1a}\zeta_{2b}}{p_2}\big(i\epsilon^{ab}\epsilon^{z_1 z_3 p_2}-(z_1\cdot p_2)\slashed{z}_3^{ab}\big).
\end{align}
After redefining
\begin{align}
    \langle J_{3/2}O_{1/2}J\rangle_{new,nh}=\langle J_{3/2}O_{1/2}J\rangle_{nh}-\langle J_{3/2}O_{1/2}J\rangle_{contact},
\end{align}
we find,
\begin{align}
    \langle J_{3/2}^{-}O_{1/2}^{-}J^{-}\rangle_{new,nh}= \langle J_{3/2}^{+}O_{1/2}^{-}J^{-}\rangle_{new,nh}=0,
\end{align}
which validates both statements (\eqref{statement2} and \eqref{statement3}).
\subsection*{\textbf{Example 4:}~$\langle T J_{3/2}O_{1/2}\rangle$}
Let us now consider an example that involves the stress tensor.
The nonhomogeneous part of this correlator is not zero in both the $(- - -)$ and $(+ - -)$ helicity sectors:
\begin{align}
    &\langle T^{-}J_{3/2}^{-}O_{1/2}^{-}\rangle_{nh}=-\frac{\langle 1 2\rangle^3\langle 3 1\rangle(p_2+p_3)}{p_1^2p_2^{3/2}\sqrt{p_3}}\notag\,,\\
    &\langle T^{+}J_{3/2}^{-}O_{1/2}^{-}\rangle_{nh}=\frac{\langle 2 3\rangle^3(E-2p_3)(E-2p_2)^3(p_2+p_3)}{\langle 1 2\rangle\langle 3 1\rangle^3p_1^2p_2^{3/2}\sqrt{p_3}}\,.
\end{align}
Consider the following contact term:
\begin{align}
   z_{1\mu}z_{1\nu} z_{2abc}\zeta_{3e}\langle T^{\mu\nu}J_{3/2}^{abc}O_{1/2}^{e}\rangle_{contact}=(p_2+p_3)\slashed{z}_1^{ab}\slashed{z}_1^{ce}z_{2abc}\zeta_{3e}.
\end{align}
By redefining
\begin{align}
    \langle T J_{3/2} O_{1/2}\rangle_{new,nh}= \langle T J_{3/2} O_{1/2}\rangle_{nh}- \langle T J_{3/2} O_{1/2}\rangle_{contact},
\end{align}
we obtain
\begin{align}
    \langle T^{-}J_{3/2}^{-}O_{1/2}^{-}\rangle_{new,nh}=\langle T^{+}J_{3/2}^{-}O_{1/2}^{-}\rangle_{new,nh}=0,
\end{align}
thereby validating statements $2$ and $3$ (\eqref{statement2} and \eqref{statement3}) once again.

To summarize, we have established in this appendix that our statements about correlators inside the triangle (\eqref{statement1}, \eqref{statement2} and \eqref{statement3}) are robust and generic.

\section{An argument for statements 2 and 3 in section \ref{sec:CFT3corrfacts} via the CFT-S matrix correspondence}\label{appendix:CFTsmatrixstuff}
In this appendix, we shall present an argument for statements $2$ and $3$ (\eqref{statement2} and \eqref{statement3}) by employing the connection between CFT correlators and scattering amplitudes in one higher dimension.

By the AdS/CFT duality, CFT correlators are dual to AdS scattering amplitudes. Thus, we can take a flat space limit to obtain the flat space scattering amplitude from the CFT correlator \cite{Maldacena:2011nz,Raju:2012zr}. This limit corresponds to taking the total energy $E=\sum_{i=1}^{n}p_i\to 0$. We shall now use this correspondence to motivate the results we obtained above.

Let us first consider $\langle TTT\rangle_{nh}$. This nonhomogeneous stress tensor three-point function is dual to the Einstein gravity graviton three-point amplitude \cite{Jain:2021qcl}. By taking the flat space limit of this correlator, the correspondence tells us that we should obtain the flat space Einstein gravity three-point amplitude. In flat space, the Einstein gravity three-point function has no contribution in the $(- - -)$ and $(+++)$ helicity configurations. For higher spins and amplitudes involving different spins as well, non-higher-derivative interactions are zero in the $(---)$ and $(+++)$ helicities, a fact that is most easily seen while attempting to write down such interactions in light-cone variables and noticing that such terms in the Lagrangian cannot simultaneously satisfy dimensional analysis, helicity neutrality, and locality. Therefore, the corresponding AdS amplitude/CFT correlators should be free of total energy ($E$) poles. It is thus plausible that such contributions can be removed via the addition of contact terms, which is indeed what we showed in section \ref{sec:CFT3corrfacts} and appendix \ref{appendix:halfintcase} for not only $\langle TTT\rangle$ but for many more examples.

Let us now discuss the zero-helicity sector. In flat space, three-point amplitudes with net zero-helicity have been shown to be inconsistent with unitarity and locality \cite{McGady:2013sga}. This was proved by showing that the zero-helicity three-point amplitudes cannot be glued together to produce a valid (local, consistent with unitary) four-point amplitude and hence such three-point seeds should vanish. While we do not have such an argument in CFT, we can make use of the CFT/S-matrix correspondence again. Assume that the CFT correlator exists in the zero-helicity sector. By taking a flat space limit, we recover a zero-helicity flat space amplitude. However, as argued above, such amplitudes do not exist and hence the CFT correlator should vanish in the flat space limit. Thus, our conclusion is that zero-helicity configurations of correlators cannot have total energy singularities. Therefore, it is plausible to remove such contributions via adding contact terms to the correlators and is exactly what we found in section \ref{sec:CFT3corrfacts} and appendix \ref{appendix:halfintcase}.

\section{Correlators outside the triangle: Examples}\label{appendix:outsidetriangleexamples}
In this appendix, we shall illustrate via three examples, the validity of statements $4$ and $5$ (\eqref{statement4} and \eqref{statement5}) presented in section \ref{sec:CFT3corrfacts}.
Let us begin with the example of $\langle J_4 J_1 J_1\rangle$ following \cite{Jain:2021whr}.
\subsection*{Example 1: $\langle J_4 J_1 J_1\rangle$}
The $FF-FB$ correlator in the independent helicity configurations takes the following forms \cite{Jain:2021whr}:
\begin{align}
    \langle J_4^{-} J_1^{-} J_1^{-}\rangle_{FF-FB}&=-\langle 12\rangle^4\langle 31\rangle^4\langle \bar{2}\bar{3}\rangle^2\frac{3 E^5+5 E^4 k_1+8 E^3 k_1^2+12 E^2 k_1^3+16 E k_1^4+16 k_1^5}{524288 E^8 k_1^4}, \nonumber\\
 \langle J_4^{+} J_1^{-} J_1^{-}\rangle_{FF-FB}&=-\langle 23\rangle^6\langle \bar{1}\bar{2}\rangle^4\langle \bar{3}\bar{1}\rangle^4\frac{3E+k_1}{524288 E^8 k_1^4},~~~ \langle J_4^{-} J_1^{-} J_1^{+}\rangle_{FF-FB}=0.
\end{align}
This statement is perfectly consistent with statement $4$ \eqref{statement4}, i.e. the vanishing of $FF-FB$ when the helicities of the spin one currents are different. Let us now examine the $FF+FB$ correlator. In spinor-helicity variables we have \cite{Jain:2021whr}
\begin{align}
    \langle J_4^{-} J_1^{-} J_1^{-}\rangle_{FB+FF}&=0,~~~\langle J_4^{+} J_1^{-} J_1^{-}\rangle_{FB+FF}=0\,,\nonumber\\
    \langle J_4^{-} J_1^{-} J_1^{+}\rangle_{FB+FF}&=\langle 12\rangle^6\langle 31\rangle^2\langle \bar{2}\bar{3}\rangle^4 \frac{5 E^3+5 E^2 k_1+4 E k_1^2+2 k_1^3}{262144 E^8 k_1^4}\,,
\end{align}
which is in perfect agreement with statement 5 \eqref{statement5} as the correlator vanishes when the helicities of the spin one currents coincide. 

Let us now proceed to an example which involves half-integer spin insertions.
\subsection*{\textbf{Example 2:} $\langle O_{1/2}O_{1/2}T\rangle$}
This correlator can be computed via Wick contractions. For the free bosonic stress tensor we have
\begin{align}
    &\zeta_{1a}\zeta_{2b}z_{3\alpha}z_{3\beta}\langle O^a_{1/2}(p_1)O^b_{1/2}(p_2)T^{\alpha\beta}_{\textbf{FB}}(p_3)\rangle=2\zeta_{1a}\zeta_{2\alpha}z_{3\alpha}z_{3\beta}(\sigma^{\rho})^{ab}\int\frac{d^3 l}{(2\pi)^3}\frac{l^\rho(l^\alpha-p_1^\alpha)(l^\beta-p_1^\beta)}{l^2(l-p_1)^2(l+p_2)^2}\notag\\
    &=-\frac{\zeta_{1a}\zeta_{2b}(p_1\cdot z_3)}{32 p_1 p_2E^3}\bigg((p_1\cdot z_3)(E+2p_2)\bigg(-p_2(\slashed{p_1}^{ab})+p_1(\slashed{p_2}^{ab})\bigg)+2p_1p_2E(E+p_3)(\slashed{z_3})^{ab}\bigg).
\end{align}
A similar computation for the free fermionic stress tensor yields,
\begin{align}
    &\zeta_{1a}\zeta_{2b}z_{3\alpha}z_{3\beta}\langle O^a_{1/2}(p_1)O^b_{1/2}(p_2)T^{\alpha\beta}_{\textbf{FF}}(p_3)\rangle=\zeta_{1a}\zeta_{2b}z_{3\alpha}z_{3\beta}(\sigma^\beta)^d_c(\sigma^{\rho})^{cb}(\sigma^\chi)^{a}_d\int\frac{d^3 l}{(2\pi)^3}\frac{(l^\alpha-p_1^\alpha)(l^\rho+p_2^\rho)(l^\chi-p_1^\chi)}{l^2(l-p_1)^2(l+p_2)^2}\notag\\
    &=\frac{\zeta_{1a}\zeta_{2b}}{32 p_1 p_2 p_3 E^3}\bigg(-(p_1\cdot z_3)^2\bigg((4(p_1+p_2)^3+(p_1+p_2)(10p_1+11p_2)p_3+3(2p_1+3p_2)p_3^2)\slashed{p_1}^{ab}-(1\leftrightarrow 2) \bigg)\notag\\
    &+(p_1\cdot z_3)(p_1+p_2)(2p_1+2p_2+3p_3)(p_1^2+p_2^2-p_3^2)(\slashed{z_3})^{ab}-2i(p_1\cdot z_3)(p_1+p_2)(2p_1+2p_2+3p_3)\epsilon^{ab}\epsilon^{z_3 p_1 p_2}\bigg).
\end{align}
\normalsize
We now consider the difference and sum of the bosonic and fermionic contributions along with the addition of a contact term:
\begin{align}
    &\langle O_{1/2}O_{1/2}T\rangle_{\textbf{FF-FB}}=\langle O_{1/2}O_{1/2}T_{\textbf{FF}}\rangle-\langle O_{1/2}O_{1/2}T_{\textbf{FB}}\rangle-\mathcal{C}_{O_{\frac{1}{2}}O_{\frac{1}{2}}T}\,,\notag\\
    &\langle O_{1/2}O_{1/2}T\rangle_{\textbf{FB+FF}}=\langle O_{1/2}O_{1/2}T_{\textbf{FB}}\rangle+\langle O_{1/2}O_{1/2}T_{\textbf{FF}}\rangle+\mathcal{C}_{O_{\frac{1}{2}}O_{\frac{1}{2}}T}\,,
\end{align}
where the contact term is given by
\begin{equation}
   \mathcal{C}_{O_{\frac{1}{2}}O_{\frac{1}{2}}T}=\frac{1}{8p_3}\zeta_{1a}\zeta_{2b}(p_1\cdot z_3)(\slashed{z_3})^{ab}\,.
\end{equation}
We now find in accordance with statement $4$ \eqref{statement4} that the $FF-FB$ correlator vanishes when the $O_{1/2}$ operators have different helicities.
\small
\begin{align}
    &\langle O_{1/2}^{-}(p_1)O_{1/2}^{-}(p_2)T^{-}(p_3)\rangle_{\textbf{FF-FB}}=-\frac{\langle 2 3\rangle^2\langle 3 1\rangle^2\bigg(4(p_1+p_2)^4+7(p_1+p_2)^3p_3-3(p_1+p_2)^2p_3^2-11(p_1+p_2)p_3^3+3p_3^4\bigg)}{128\langle 1 2\rangle\sqrt{p_1}\sqrt{p_2}p_3^3 E^3},\notag\\
    &\langle O_{1/2}^{-}(p_1)O_{1/2}^{+}(p_2)T^{-}(p_3)\rangle_{\textbf{FF-FB}}=0,\notag\\
    &\langle O_{1/2}^{+}(p_1)O_{1/2}^{-}(p_2)T^{-}(p_3)\rangle_{\textbf{FF-FB}}=0,\notag\\
    &\langle O_{1/2}^{-}(p_1)O_{1/2}^{-}(p_2)T^{+}(p_3)\rangle_{\textbf{FF-FB}}=-\frac{\langle 1 2\rangle^3(4(p_1+p_2)+3p_3)((p_1-p_2)^2-p_3^2)^2}{128\langle 2 3\rangle^2\langle 3 1\rangle^2\sqrt{p_1}\sqrt{p_2}p_3^3}.
\end{align}
We also see that the FF+FB correlator obeys statement $5$ \eqref{statement5}:
\begin{align}
    &\langle O_{1/2}^{-}(p_1)O_{1/2}^{-}(p_2)T^{-}(p_3)\rangle_{\textbf{FB+FF}}=0,\notag\\
    &\langle O_{1/2}^{-}(p_1)O_{1/2}^{+}(p_2)T^{-}(p_3)\rangle_{\textbf{FB-FF}}=-\frac{\langle 2 3\rangle\langle 3 1\rangle^3(p_1+p_2)(p_1+p_2-p_3)^2(2p_1+2p_2+3p_3)}{64\langle 1 2\rangle^2\sqrt{p_1}\sqrt{p_2}p_3^3E^2},\notag\\
    &\langle O_{1/2}^{+}(p_1)O_{1/2}^{-}(p_2)T^{-}(p_3)\rangle_{\textbf{FB-FF}}=\frac{\langle 2 3\rangle^3\langle 3 1\rangle(p_1+p_2)(p_1+p_2-p_3)^2(2p_1+2p_2+3p_3)}{64\langle 1 2\rangle^2\sqrt{p_1}\sqrt{p_2}p_3^3 E^2},\notag\\
    &\langle O_{1/2}^{-}(p_1)O_{1/2}^{-}(p_2)T^{+}(p_3)\rangle_{\textbf{FB-FF}}=0.
\end{align}
To conclude this appendix, let us consider one more example, this time with a correlator involving a spin $4$ current and two spin half operators.
\subsection*{\textbf{Example~3}$:\langle J_4 O_{1/2}O_{1/2}\rangle$}
Computing this correlator via Wick contractions, we see that the $FF-FB$ and $FF+FB$ combinations automatically obey statements $5$ and $6$ (\eqref{statement5} and \eqref{statement6}). The expressions in the independent helicities are as follows:
\begin{align}
    &\langle J_4^{-}O_{1/2}^{-}O_{1/2}^{-}\rangle_{FF-FB}=\frac{\langle 1 2\rangle^4 \langle 3 1\rangle^4}{4096 \langle 2 3\rangle^3 E^5 p_1^4\sqrt{p_2 p_3}}\bigg(15 E^7-70 E^6 p_1+84 E^5 p_1^2-128 p_1^7\bigg)\,,\notag\\
    &\langle J_4^{-}O_{1/2}^{+}O_{1/2}^{+}\rangle_{FF-FB}=-\frac{\langle 12 \rangle^4\langle 3 1\rangle^4}{4096 \langle 2 3\rangle^5 E^3 p_1^4\sqrt{p_2 p_3}}(E-2p_1)^5\bigg(15 E^2+10 Ep_1+4p_1^2\bigg)\,,\notag\\
    &\langle J_4^{-}O_{1/2}^{\pm}O_{1/2}^{\mp}\rangle_{FF-FB}=0\,,
\end{align}
and
\begin{align}
    &\langle J_4^{-}O_{1/2}^{-}O_{1/2}^{+}\rangle_{FF+FB}=\frac{\langle 12\rangle^5\langle 3 1\rangle^3}{1024 \langle 2 3\rangle^4 E^4 p_1^4\sqrt{p_2p_3}}(E-2p_1)^4\bigg(5 E^3+5 E^2 p_1+4 p_1^2+2p_1^3\bigg)\,,\notag\\
    &\langle J_4^{-}O_{1/2}^{+}O_{1/2}^{+}\rangle=0\,,\notag\\
    &\langle J_{4}^{-}O_{1/2}^{-}O_{1/2}^{-}\rangle=0\,,
\end{align}
thereby verifying the robustness of our statements.

\section{The correct parity even vs. parity odd relation outside the triangle}\label{appendix:epToutsidetriangle}
In section \ref{sec:CFT3corrfacts}, we made a statement \eqref{statement6} about the correct prescription for the epsilon transform for correlators in slightly-broken higher-spin theories whose insertions violate the spin triangle inequality. In this appendix, we prove said statement by employing slightly-broken higher-spin equations in the quasi fermionic theory discussed in section \ref{sec:CS+fermions} (It is straightforward to repeat such an analysis for the quasi bosonic case of appendix \ref{appendix:QBchirallimit}).\\

Without loss of generality assume that $s_1>s_2+s_3$, $s_2\ge s_3$ for a correlator $\langle J_{s_1}J_{s_2}J_{s_3}\rangle$ that is outside the triangle. It is easy to show that in this correlator, $J_{s_2}, J_{s_3}$ behave like exactly conserved currents and the nonconservation is entirely due to $J_{s_1}$.\footnote{This is easy to see as follows: Assume on the contrary that $\partial\cdot J_{s_2}\sim J_{s_1}\partial^n J_{s_3}$ (schematically). Then the LHS has scaling dimension $s_2+2+\order{\frac{1}{N}}$ while the RHS has $s_1+s_3+n+2$. Matching these we obtain $s_2=s_1+s_3+n$. Since $n\ge 0$, this is a contradiction and hence $\partial\cdot J_{s_2}$ does not contain $J_{s_1}\partial^n J_{s_3}$. Note that this is similar to the statement that the only Ward-Takahashi identity for correlators outside the triangle is due to the highest spin. } An example of such a correlator is $\langle J_4 J_1 J_1\rangle$. It was observed in \cite{Jain:2021whr} that for $\langle J_4 J_1 J_1\rangle$, the epsilon transformation to obtain the odd piece should be performed with respect to the $J_1$ operators, i.e. the currents that are exactly conserved in this correlator. Does this pattern extend to any correlator outside the triangle? We shall answer this question via the analysis of the slightly-broken higher-spin equations. Before we present the detailed analysis, we summarize our results: the parity odd part of the correlator $\langle J_{s_1}J_{s_2}J_{s_3}\rangle$ (with $s_1>s_2+s_3$, $s_2\ge s_3$) can be obtained in the following equivalent ways: 
 \begin{align}\label{thedifferentEPT}
&\langle J_{s_1}J_{s_2}J_{s_3}\rangle_{odd}=\langle J_{s_1}\epsilon\cdot J_{s_2}J_{s_3}\rangle_{FF-FB}=\langle J_{s_1}J_{s_2}\epsilon\cdot J_{s_3}\rangle_{FF-FB}\,,
 \end{align}
 which are epsilon transforms with respect to $J_{s_2}$ or $J_{s_3}$, the currents that are conserved in this correlator. An alternative quick way to arrive at the same conclusion is to use the fact that the epsilon transform is conformally invariant if and only if it is performed on a conserved current (see \cite{Caron-Huot:2021kjy}) and in the above case, this picks out the ones with respect to $J_{s_2}$ and $J_{s_3}$.
 
 Further validation of \eqref{thedifferentEPT} comes from the fact that it has the following implications on the $FF-FB$ correlators ($h_1$ is the helicity of $J_{s_1}$ below):
 \begin{align}
     \langle J_{s_1}^{h_1}J_{s_2}^{\pm}J_{s_3}^{\mp}\rangle_{FF-FB}=0\,,
 \end{align}
which is nothing but statement $4$ \eqref{statement4}. Let us now first illustrate how to obtain \eqref{thedifferentEPT} via an example:
\subsection*{\textbf{An Example:}~$\langle J_5 T J\rangle$}
Consider the equation:
\begin{align}
    &\langle[ Q_{3}^{\mu\nu},J_5^{\rho\sigma\gamma\omega\phi}(x_1)]T^{\alpha\beta}(x_2)O_2(x_3)\rangle+\langle J_5^{\rho\sigma\gamma\omega\phi}(x_1)[ Q_{3}^{\mu\nu},T^{\alpha\beta}(x_2)]O_2(x_3)\rangle+\langle J_5^{\rho\sigma\gamma\omega\phi}(x_1)T^{\alpha\beta}(x_2)[Q_{3}^{\mu\nu},O_2(x_3)]\rangle\notag\\
    &=\int d^3 x \langle \partial_{\psi}J_3^{\mu\nu\psi}(x)J_5^{\rho\sigma\gamma\omega\phi}(x_1)T^{\alpha\beta}(x_2)O_2(x_3)\rangle\,,
\end{align}
where $Q_3$ is a higher-spin charge corresponding to the current $J_3$ and the other operators have their usual meaning.
 We then use the current algebra, the nonconservation of the spin-$3$ current, convert it to the momentum space, and substitute the expansion of the correlators appearing in the resulting equation.\footnote{The procedure for the same follows the analysis of \cite{Jain:2022ajd}. The nonconservation operator of the $J_3$ current is also given in (B.15) in \cite{Jain:2022ajd}.} By expanding the higher-spin equation about $\Tilde{\lambda}=i$ we obtain the following equation that has to be satisfied:
 \begin{align}\label{JepT}
     \langle J_5^{\rho\sigma\gamma\omega\phi}T^{\alpha\beta}J^{\mu}\rangle_{odd}=\frac{\epsilon^{\mu a b}p_{3a}}{p_3}\big(\langle J_5^{\rho\sigma\omega\gamma\phi}T^{\alpha\beta}J^{b}\rangle_{FF}-\langle J_5^{\rho\sigma\omega\gamma\phi}T^{\alpha\beta}J^{b}\rangle_{FB}\big)\,,
 \end{align}
 which is the epsilon transform with respect to $J$.

 Let us now show that by solving a different higher-spin equation we obtain an epsilon transform with respect to $T$. Consider,
 \begin{align}
     &\langle[Q_4^{\nu\alpha\beta},J_5^{\rho\sigma\gamma\omega\phi}(x_1)]O_2(x_2)J^\mu(x_3)\rangle+\langle J_5^{\rho\sigma\gamma\omega\phi}(x_1)[Q_4^{\nu\alpha\beta},O_2(x_2)]J^\mu(x_3)\rangle+\langle J_5^{\rho\sigma\gamma\omega\phi}(x_1)O_2(x_2)[Q_4^{\nu\alpha\beta},J^\mu(x_3)]\rangle\notag\\
     &=\int d^3 x \langle \partial_\psi J_4^{\psi\nu\alpha\beta}(x)J_5^{\rho\sigma\gamma\omega\phi}(x_1)O_2(x_2)J^\mu(x_3)\rangle\,.
 \end{align}
 Similar to the previous exercise, we use the higher-spin algebra, the nonconservation equation, convert to Fourier space and use the expansion of the correlators. We then expand the resulting equation about $\Tilde{\lambda}=i$ which yields the following equation:
 \begin{align}\label{TepT}
     \langle J_5^{\rho\sigma\gamma\omega\phi}T^{\alpha\beta}J^{\mu}\rangle_{odd}=\frac{\epsilon^{\alpha a b}p_{2a}}{p_2}\big(\langle J_5^{\rho\sigma\omega\gamma\phi}T^{b\beta}J^{\mu}\rangle_{FF}-\langle J_5^{\rho\sigma\omega\gamma\phi}T^{b\beta}J^{\mu}\rangle_{FB}\big)\,,
 \end{align}
 which is an epsilon transform with respect to $T$.
 
 Since we now have two different expressions for the odd part of the correlator, viz. \eqref{JepT} and \eqref{TepT}, let us contract these expressions with the appropriate polarization vectors and then equate them to derive the consequences. In the $(\pm - -)$ and $(\pm + +)$ helicity configurations, that is, when $T$ and $J$ have the same helicities, it is easy to see that \eqref{JepT} and \eqref{TepT} agree. However, when $T$ and $J$ have opposite helicities, \eqref{JepT} yields
 \begin{align}
     \langle J_5^{\pm}T^{-}J^{+}\rangle_{odd}=i\langle J_5^{\pm}T^{-}J^{+}\rangle_{FF-FB}\,,
 \end{align}
 whereas \eqref{TepT} gives
 \begin{align}
     \langle J_5^{\pm}T^{-}J^{+}\rangle_{odd}=-i\langle J_5^{\pm}T^{-}J^{+}\rangle_{FF-FB}.
 \end{align}
 For these two expressions to be consistent we would require,
 \begin{align}
     &\langle J_5^{\pm}T^{-}J^{+}\rangle_{FF-FB}=0\,.
 \end{align}
 By similar considerations, we also see that we have to require
 \begin{align}
     &\langle J_5^{\pm}T^{+}J^{-}\rangle_{FF-FB}=0\,.
 \end{align}
 Thus, the free fermion minus critical boson correlator should vanish whenever $T$ and $J$ have opposite helicities which is nothing but a particular case of statement $4$ \eqref{statement4}! Let us now generalize these findings to more general correlators.
 \subsection*{Generalization to arbitrary spins}
 If we repeat the previous analysis for say $\langle J_4 J_1 J_1\rangle$ we find
 \begin{align}
     \langle J_4^{h_1}J^{\mp}J^{\pm}\rangle_{FF-FB}=0\,,
 \end{align}
 which is indeed the case, see \cite{Jain:2021whr}. 
 One issue to be wary of in these explicit calculations are contact terms that have to be carefully removed from the correlators to obtain these results much like we did in section \ref{sec:CFT3corrfacts} for correlators that are inside the triangle.\\
 For arbitrary correlators with $s_1>s_2+s_3$, $s_2\ge s_3$, we see that the slightly-broken higher-spin equations always yield epsilon transforms with respect to either of the lower two spins. Consistency of these formulae demands:
 \begin{align}
     \langle J_{s_1}^{h_1}J_{s_2}^{\pm}J_{s_3}^{\mp}\rangle_{FF-FB}=0\,.
 \end{align}
Let us now prove this. Consider the following slightly-broken higher-spin equation (assuming that $s_1>s_2+s_3$, $s_2>s_3$. We also suppress the indices of the operators to simplify notation)
\begin{align}\label{s1s2s3sbhse}
    \langle [Q_{s_3+2}^{\mu_1\cdots \mu_{s_3+1}},J_{s_1}J_{s_2}O_2]\rangle=\int d^3 x \langle\partial_{\mu_{s_3+2}}J_{s_3+2}^{\mu_1\cdots\mu_{s_3+2}}J_{s_1}J_{s_2}O_2\rangle.
\end{align}
Schematically, for the commutators on the LHS, we have,
\begin{align}\label{s1s2s3com}
&[Q_{s_3+2}^{\mu_1\cdots \mu_{s_3+1}},J_{s_1}]=J_{s_1+s_3}+\cdots ,\notag\\
&[Q_{s_3+2}^{\mu_1\cdots\mu_{s_3+1}},O_2]=\epsilon^{\mu_1 a b}\big(\partial^{\mu_2}\partial^a J_{s_3}^{b\mu_3\cdots \mu_{s_3+1}}+\cdots+\partial^{\mu_2}\partial^{\mu_3}\partial^{\mu_4}\partial^a J_{s_3-2}^{b\mu_5\cdots \mu_{s_3+1}}+\cdots\big),
\end{align}
while for the non conservation of the spin $s_3+2$ current on the RHS we have\footnote{Note that these are the only terms that contribute after the large-$N$ factorization of the five-point correlator in the RHS of \eqref{s1s2s3sbhse} since $s_3<s_2,s_1$ and hence $J_{s_1}$ and $J_{s_2}$ do not appear in the non conservation of $J_{s_3}$ by dimensional analysis.}
\begin{align}\label{s1s2s3noncons}
    \partial_{\mu_{s_3}+2}J_{s_3+2}^{\mu_1\cdots \mu_{s_3+2}}=\frac{\Tilde{\lambda}}{\Tilde{N}(1+\Tilde{\lambda}^2)}\big(O_2 \partial^{\mu_2}J_{s_3}^{\mu_1\cdots \mu_{s_3+1}}+\cdots+O_2\partial^{\mu_2}\partial^{\mu_3}\partial^{\mu_4}J_{s_3-2}^{\mu_1\mu_5\cdots\mu_{s_3+1}}+\cdots\big).
\end{align}
We now use \eqref{s1s2s3com} and \eqref{s1s2s3noncons} in \eqref{s1s2s3sbhse}, perform large N factorization on the RHS, convert the resulting equation to momentum space obtaining:
\small
\begin{align}\label{sbhses1s2s3}
    &\big(\cdots+\epsilon^{\mu_1 ab}p_3^{a}p_3^{\mu_2}\langle J_{s_1}J_{s_2}J_{s_3}^{b\mu_3\cdots \mu_{s_3+1}}\rangle+\cdots+\epsilon^{\mu_1 ab}p_3^{a}p_{3}^{\mu_2}p_3^{\mu_3}p_3^{\mu_4}\langle J_{s_1}J_{s_2}J_{s_3-2}^{b\mu_5\cdots \mu_{s_3+1}}\rangle+\cdots\big)\notag\\
    &=\frac{\Tilde{\lambda}}{\Tilde{N}(1+\Tilde{\lambda}^2)}\big(\langle O_2(p_3)O_2(-p_3)\rangle p_3^{\mu_2}\langle J_{s_1}J_{s_2}J_{s_3}^{\mu_1\cdots \mu_{s_3+1}}\rangle+\cdots+\langle O_2(p_3)O_2(-p_3)\rangle p_3^{\mu_2}p_3^{\mu_3}p_3^{\mu_4}\langle J_{s_1}J_{s_2}J_{s_3-2}^{\mu_1\mu_5\cdots \mu_{s_3+1}}\rangle+\cdots\big)
\end{align}
\normalsize
where we kept explicit only terms involving three-point correlators with all three insertions having non-zero spin. We now use the expansion of these correlators provided in \eqref{s1s2s3outsidetriangle} and \eqref{nfnoddnbcoeffs}. We then expand the equation about the point $\Tilde{\lambda}=i$ and analyze the resulting pole equation. Consider now the $s_3=2$ case. The resulting pole equation \eqref{sbhses1s2s3} is solved if
\begin{align}\label{eptT}
    \langle J_{s_1}J_{s_2}T^{\mu_1\mu_2}\rangle_{odd}=\frac{\epsilon^{\mu_1 a b}p_3^{a}}{p_3}\langle J_{s_1}J_{s_2}T^{b\mu_2}\rangle_{FF-CB}.
\end{align}
For the $s_3=4$ case, we obtain (upon using \eqref{eptT})
\begin{align}\label{eptJ4}
    \langle J_{s_1}J_{s_2}J_4^{\mu_1\mu_2\mu_3\mu_4}\rangle_{odd}=\frac{\epsilon^{\mu_1 a b}p_3^{a}}{p_3}\langle J_{s_1}J_{s_2}J_4^{b\mu_2\mu_3\mu_4}\rangle_{FF-CB}.
\end{align}
Using \eqref{eptT} and \eqref{eptJ4} in the $s_3=6$ case yields an epsilon transform with respect to $J_6$. Continuing in this fashion and using the previous results recursively shows that,
\begin{align}
    \langle J_{s_1}J_{s_2}J_{s_3}^{\mu_1\mu_2\mu_3\mu_4\cdots\mu_{s_3}}\rangle_{odd}=\frac{\epsilon^{\mu_1 a b}p_3^{a}}{p_3}\langle J_{s_1}J_{s_2}J_{s_3}^{b\mu_2\mu_3\mu_4\cdots \mu_{s_3}}\rangle_{FF-CB}~\forall s_1>s_2+s_3,s_2>s_3
\end{align}
One can show by similar methods that we can also obtain an epsilon transform with respect to $J_{s_3}$ even if $s_3>s_2$, that is, if $J_{s_3}$ is the second lowest spin in the correlator. This generalizes the epsilon transform with respect to $T$ \eqref{TepT} in the example of $\langle J_5 TJ\rangle$\footnote{The only complication in this case is that the non conservation of \eqref{s1s2s3noncons} can contain a $J_{s_2}$. However, the commutator of $Q_{s_3+2}$ with $O_2$ on the LHS of \eqref{s1s2s3sbhse} will also contain a $J_{s_2}$ and these terms will cancel out.}.

\section{Building four-point functions using Conformal partial waves}\label{appendix:CPW}
In this appendix, we use conformal partial waves to construct several four-point functions purely from the anti-chiral vertices. 

The conformal partial wave (CPW) $W_s$ that represents the contribution to a correlator due to the exchange of a current with spin $s$ and scaling dimension $s+1$ is given by (in the $s$ channel) \cite{Simmons-Duffin:2012juh}
\begin{align}\notag
    W_s^{(s)}=\sum_{z}\int \frac{d^3 x_5 d^3 x_6}{|x_5-x_6|^{2(2-s)}}\langle J_{s_1}(x_1,z_1)J_{s_2}(x_2,z_2)J_s(x_5,z)\rangle\langle J_s(x_6,z)J_{s_3}(x_3,z_3)J_{s_4}(x_4,z_4)\rangle.
\end{align}
Upon obtaining the momentum space counterpart to the above formula via the Fourier transform and choosing to work in the helicity basis, we obtain the following result:
\begin{align}\notag
    W_s^{(s)}=\sum_{h=+,-}\frac{1}{|p_1+p_2|^{2s-1}}\langle J_{s_1}^{h_1}(p_1)J_{s_2}^{h_2}(p_2)J_s^{h}(-p_1-p_2)\rangle\langle J_{s_3}^{h_3}(p_3)J_{s_4}^{h_4}(p_4)J_s^{-h}(-p_3-p_4)\rangle.
\end{align}
Similarly, one obtains the CPW corresponding to a $\Delta=2$ scalar exchange,
\begin{align}
    W_{\Delta=2}^{(s)}=\frac{1}{|p_1+p_2|}\langle J_{s_1}^{h_1}(p_1)J_{s_2}^{h_2}(p_2)O_2(-p_1-p_2)\rangle\langle J_{s_3}^{h_3}(p_3)J_{s_4}^{h_4}(p_4)O_2(-p_3-p_4)\rangle.
\end{align}
To obtain a crossing symmetric result, we add to the $s-$channel CPWs, the $t$ and $u$ channel contributions.\\ 
In principle, one has to take into account double-trace exchanges as well. However, in slightly-broken higher-spin theories, at least for four-point correlators where all operators are spinning, the four-point functions are given purely in terms of the free fermionic and free bosonic theory results plus $O_2$ conformal blocks where possible \cite{Jain:2023juk}. Since the free theory correlators are given purely in terms of single-trace exchange twist conformal blocks \cite{Sleight:2017pcz,Jain:2023juk}, there are no additional double trace contributions that we need to add to the result obtained after summing up all the exchanges. Since we work with the three-point functions that are purely anti-chiral (they have support only in configurations with net negative helicity), a given four-point function will get contributions only from a finite number of exchanges.

The discussion that is to follow is a parallel to the bulk analysis of section \ref{sec:holo} from the perspective of the boundary CFT. 
\subsection*{\textbf{Example 1:}~$\langle \Tilde{O}_2 \Tilde{O}_2 \Tilde{O}_2 \Tilde{O}_2\rangle$}
This correlator cannot receive a contribution from scalar exchanges since $\langle \Tilde{O}_2\Tilde{O}_2\Tilde{O}_2\rangle=0$. The CPW for a spin $s$ exchange is of the form $\langle \Tilde{O}_2 \Tilde{O}_2 \Tilde{J}_s^{-}\rangle \langle \Tilde{O}_2 \Tilde{O}_2 \Tilde{J}_{s}^{+}\rangle$ which is also zero since $\langle \Tilde{O}_2 \Tilde{O}_2 \Tilde{J}_s^{+}\rangle=0$ in the anti-chiral limit. Therefore, in this limit, this four-point correlator is identically zero.
\subsection*{\textbf{Example 2:}~$\langle \Tilde{J}_s \Tilde{O}_2 \Tilde{O}_2\Tilde{O}_2\rangle$}
Since we are working with currents that are singlets of the gauge group, this correlator is nonzero if and only if the spin $s$ is even. 

Let us thus first consider the $s=2$ case, i.e. $\langle \Tilde{T}\Tilde{O}_2\Tilde{O}_2\Tilde{O}_2\rangle$. When the stress tensor has positive helicity, it is obvious that this correlator vanishes. When it has a negative helicity, the only possible contribution is of the form, $\langle \Tilde{T}^{-}\Tilde{J}^{+}\Tilde{O}_2\rangle\langle \Tilde{O}_2 \Tilde{O}_2 \Tilde{J}^{-}\rangle$. However, both the terms in this product are identically zero since the current $\Tilde{J}$ and the scalars $\Tilde{O}_2$ are singlets.

For higher-spins, the story is a little different. For instance, for $s=4$ there exists a nonzero contribution, viz.
\begin{align}
    \langle \Tilde{J}_4^{-}\Tilde{O}_2\Tilde{O}_2\Tilde{O}_2\rangle\sim \frac{1}{|p_1+p_2|^{3}}\langle \Tilde{J}_4^{-}\Tilde{O}_2 \Tilde{T}^{+}\rangle\langle \Tilde{O}_2 \Tilde{O}_2 \Tilde{T}^{-}\rangle+\text{t and u channel terms}.
\end{align}
Similarly, it is easy to show that for spin $s$, there are contributions due to exchanges of spins from $2$ to $s-2$. Note however that our result \eqref{JsO2O2O2aclimit} is agnostic about the value of the spin, $s$. For spin-two, the authors of \cite{Jain:2022ajd} observed that one solution to the higher-spin equations is the critical bosonic $\langle TO_2O_2O_2\rangle$ being given as the epsilon transform of its free fermionic counterpart. If this is the case, \eqref{JsO2O2O2aclimit} would indeed become zero for $s=2$. However, based on the analysis in \cite{Jain:2022ajd}, a similar result should follow for the higher-spins as well rendering $\langle \Tilde{J}_s^{-}\Tilde{O}_2\Tilde{O}_2\Tilde{O}_2\rangle$ zero for all spins. However, as we just saw above and in section \ref{sec:beyond}, there are possible nonzero contributions to this correlator. We leave a more detailed investigation of this apparent discrepancy to the future. 

\subsection*{\textbf{Example 3:}~$\langle \Tilde{J}_s \Tilde{J}_s \Tilde{O}_2 \Tilde{O}_2\rangle$}
In the $s=1$ case, it is easy to see that this correlator receives no contributions from any exchange in all helicity configurations. The same result also holds for the $s=2$ case. For spin $3$ onwards, there can be nonzero contributions in the $(-~-~0~0)$ helicity configuration, for instance, $\langle \Tilde{J}_{3}^{-}\Tilde{O}_2 \Tilde{J}^{+}\rangle \langle \Tilde{J}_{3}^{-}\Tilde{O}_2 \Tilde{J}^{-}\rangle$ for the correlator $\langle \Tilde{J}_3^{-}\Tilde{J}_3^{-}\Tilde{O}_2\Tilde{O}_2\rangle$.

\subsection*{\textbf{Example 4:}~$\langle \Tilde{J}\Tilde{J}\Tilde{J}\Tilde{J}\rangle$}
Finally, let us look at one more four-point example with all nonzero spin insertions.
It is obvious that this correlator should vanish in the following configurations (and their permutations): $(++++),(+++-),(++--)$ and $(---+)$. The only nonzero configuration is $(----)$. Indeed, we have 
\begin{align}
\begin{aligned}
    \langle \Tilde{J}^{-}(p_1)&\Tilde{J}^{-}(p_2)\Tilde{J}^{-}(p_3)\Tilde{J}^{-}(p_4)\rangle\sim\sum_{i=s,t,u}W_{\Delta=2}^{(i)}=\\
    &=\frac{1}{|p_1+p_2|}\langle \Tilde{J}^{-}(p_1)\Tilde{J}^{-}(p_2)\Tilde{O}_2\rangle\langle \Tilde{J}^{-}(p_3)\Tilde{J}^{-}(p_4)\Tilde{O}_2\rangle+\text{t and u channels}\,.
\end{aligned}
\end{align}
Using the explicit expression of the three-point functions appearing in the RHS of the above formula we obtain (we define $s=|p_1+p_2|$ below)
\begin{align}\label{JJJJviaCPW}
    \langle \Tilde{J}^{-}(p_1)\Tilde{J}^{-}(p_2)\Tilde{J}^{-}(p_3)\Tilde{J}^{-}(p_4)\rangle\sim \frac{\langle 1 2\rangle^2\langle 3 4\rangle^2}{s(p_1+p_2+s)^2(p_3+p_4+s)^2}+(2\leftrightarrow 3)+(2\leftrightarrow 4).
\end{align}
As \eqref{JJJJviaCPW} is free of total energy singularities, it vanishes in the flat space limit, a fact that is consistent with the known results \cite{Ponomarev:2016lrm,Skvortsov:2018jea}.\\

Let us now compare and contrast \eqref{JJJJviaCPW} with the expression \eqref{Js4pointmmmm} that we obtain by first computing the four-point correlator and then taking the anti-chiral limit. In \eqref{JJJJviaCPW}, the result is simple and finite. \eqref{Js4pointmmmm} on the other hand has parts that diverge as we take the anti-chiral limit. This indicates that one really has to first take the anti-chiral limit and then compute observables rather than attempt to obtain the anti-chiral sector as a limiting case of the general results. This is a manifestation of the fact that taking an infinite sum (the contributions of all exchanged operators to a correlator) and taking a limit (the anti-chiral limit) do not commute. It will be interesting to explore this further in the future.

\section{More on four-point correlators}\label{appendix:ExtraFourPoint}
In this appendix, just as we did in section \ref{sec:ChiralHSfromCS}, we shall attempt to take the anti-chiral limit \eqref{anti-chirallimit} for a couple more four-point functions. In the examples we considered there, we saw a nice separation between the chiral and anti-chiral correlators. However, as we shall show with the two examples in this appendix, the story is more involved for more complicated correlators.
\subsection*{\textbf{Example 1:}~$\langle J_{s_1}J_{s_2}O_2O_2\rangle$}
This correlator is given  by \cite{Jain:2022ajd} (we assume that $s_1>s_2$)
\begin{align}\notag
    \langle J_{s_1}J_{s_2}O_2O_2\rangle=\frac{1}{\Tilde{N}}\bigg(\langle J_{s_1}J_{s_2}O_2O_2\rangle_{FF}+\Tilde{\lambda}\langle J_{s_1}\epsilon\cdot J_{s_2}O_2O_2\rangle_{FF-CB}+\Tilde{\lambda}^2 \langle J_{s_1}J_{s_2}O_2O_2\rangle_{CB}\bigg).
\end{align}
The four different helicity configurations of this correlator are given by
\begin{align}
    &\langle \Tilde{J}_{s_1}^{-}\Tilde{J}_{s_2}^{-}\Tilde{O}_2\Tilde{O}_2\rangle=\frac{1}{2\Tilde{N}}\bigg(\langle J_{s_1}^{-}J_{s_2}^{-}O_2O_2\rangle_{FF-CB}+e^{2i\theta}\langle J_{s_1}^{-}J_{s_2}^{-}O_2O_2\rangle_{FF+CB}\bigg),\notag\\
    &\langle \Tilde{J}_{s_1}^{-}\Tilde{J}_{s_2}^{+}\Tilde{O}_2\Tilde{O}_2\rangle=\frac{1}{2\Tilde{N}}\bigg(e^{2i\theta}\langle J_{s_1}^{-}J_{s_2}^{+}O_2O_2\rangle_{FF-CB}+\langle J_{s_1}^{-}J_{s_2}^{+}O_2O_2\rangle_{FF+CB}\bigg),\notag\\
    &\langle \Tilde{J}_{s_1}^{+}\Tilde{J}_{s_2}^{-}\Tilde{O}_2\Tilde{O}_2\rangle=\frac{1}{2\Tilde{N}}\bigg(e^{-2i\theta}\langle J_{s_1}^{-}J_{s_2}^{+}O_2O_2\rangle_{FF-CB}+\langle J_{s_1}^{-}J_{s_2}^{+}O_2O_2\rangle_{FF+CB}\bigg),\notag\\
    &\langle \Tilde{J}_{s_1}^{+}\Tilde{J}_{s_2}^{+}\Tilde{O}_2\Tilde{O}_2\rangle=\frac{1}{2\Tilde{N}}\bigg(\langle J_{s_1}^{+}J_{s_2}^{+}O_2O_2\rangle_{FF-CB}+e^{-2i\theta}\langle J_{s_1}^{+}J_{s_2}^{+}O_2O_2\rangle_{FF+CB}\bigg).
\end{align}
By taking the anti-chiral limit \eqref{anti-chirallimit}, we see that we obtain just the anti-chiral sector:
\begin{align}\label{Js1Js2O2O2example}
    &\langle \Tilde{J}_{s_1}^{-}\Tilde{J}_{s_2}^{-}\Tilde{O}_2\Tilde{O}_2\rangle=\frac{g_{ac}^2}{2}\langle J_{s_1}^{-}J_{s_2}^{-}O_2O_2\rangle_{FF+CB},\notag\\
    &\langle \Tilde{J}_{s_1}^{-}\Tilde{J}_{s_2}^{+}\Tilde{O}_2\Tilde{O}_2\rangle=\frac{g_{ac}^2}{2}e^{2i\theta}\langle J_{s_1}^{-}J_{s_2}^{+}O_2O_2\rangle_{FF-CB},\notag\\
    &\langle \Tilde{J}_{s_1}^{+}\Tilde{J}_{s_2}^{-}\Tilde{O}_2\Tilde{O}_2\rangle=0,\notag\\
    &\langle \Tilde{J}_{s_1}^{+}\Tilde{J}_{s_2}^{+}\Tilde{O}_2\Tilde{O}_2\rangle=0.
\end{align}
If we repeat the same analysis for the $s_1=s_2$ case, we obtain the same result as \eqref{Js1Js2O2O2example}. This however, poses an immediate puzzle. The second line of \eqref{Js1Js2O2O2example} would be a net-zero helicity correlator and thus not naturally belong to either the chiral or anti-chiral sectors. Indeed, the complete Chern-Simons matter theory is not a union of the (anti-)chiral sectors and contains nonchiral structures as well.

Let us now turn to the four-point correlator of a spin-$s$ current where such problems will be more pronounced.
\subsection*{\textbf{Example 2:}~$\langle  J_s J_sJ_sJ_s\rangle $}\label{appendix:JJJJ}
Consider the four-point function of the operator $J_s$. It's expression in the $(----)$ helicity configuration is given by \cite{Jain:2022ajd}
\begin{align}\label{Js4pointmmmm}
&\langle\Tilde{J}_s^{-}\Tilde{J}_s^{-}\Tilde{J}_s^{-}\Tilde{J}_s^{-}\rangle=\frac{g^2}{2}\bigg(\langle J_s^{-}J_s^{-}J_s^{-}J_s^{-}\rangle_{FF-CB}+e^{2i\theta}\langle J_s^{-}J_s^{-}J_s^{-}J_s^{-}\rangle_{FF+CB}\bigg).
\end{align}
Thus, taking the anti-chiral limit \eqref{anti-chirallimit} would cause this correlator to diverge due to the second term in \eqref{Js4pointmmmm}.\footnote{In section \ref{sec:CFT3corrfacts}, we showed that at the level of three-points, the $FF+CB$ correlators can be set to zero when the helicities of all the operators coincide via the addition of contact terms. At the four-point level, we do not yet have such a statement.} We now consider another helicity configuration, in particular, one that has net helicity zero
\begin{align}\label{Js4pointmmpp}
    \langle \Tilde{J}_s^{-}\Tilde{J}_s^{-}\Tilde{J}_s^{+}\Tilde{J}_s^{+}\rangle=\frac{g^2}{4}\bigg(\langle J_s^{-}J_{s}^{-}J_s^{+}J_s^{+}\rangle_{FF-CB}+2e^{-2i\theta}\langle J_s^{-}J_{s}^{-}J_s^{+}J_s^{+}\rangle_{FF+CB}+e^{-4i\theta}\langle J_s^{-}J_{s}^{-}J_s^{+}J_s^{+}\rangle_{FF-CB}\bigg).
\end{align}
Taking the anti-chiral limit \eqref{anti-chirallimit} we are left with the first term of \eqref{Js4pointmmpp}. Thus, we see that we obtain a zero-helicity correlator that belongs to neither the chiral or anti-chiral sectors.\footnote{At the level of three-point functions, we were able to show that correlators in the zero-helicity sector can identically be made zero via the addition of contact terms. However, we do not yet have such a statement at the level of four-points.} Thus we see that the story at four-points and beyond requires more careful considerations which we leave for future work.

\footnotesize
\bibliographystyle{apsrev4-1}
\bibliography{megabib}

\end{document}